\documentclass[default,iicol]{sn-jnl}% Default with double column layout

\jyear{2023}%

\theoremstyle{thmstyleone}%
\theoremstyle{thmstyletwo}%

\theoremstyle{thmstylethree}%

\usepackage[normalem]{ulem} % for strikethrough text

\usepackage{graphicx}
\usepackage{subfigure}

\usepackage{wrapfig}

\newcommand{\rr}[1]{#1} % to remove the track changes
\newcommand{\rdelete}[1]{} % to remove the deleted text

\begin{document}

%%%%%%%%%%%%%%%%%%%%%%%%

\title[Article Title]{Exploring Variational Auto-Encoder Architectures, Configurations, and Datasets for Generative Music Explainable AI}

%%=============================================================%%
%% Prefix   -> \pfx{Dr}
%% GivenName    -> \fnm{Joergen W.}
%% Particle -> \spfx{van der} -> surname prefix
%% FamilyName   -> \sur{Ploeg}
%% Suffix   -> \sfx{IV}
%% NatureName   -> \tanm{Poet Laureate} -> Title after name
%% Degrees  -> \dgr{MSc, PhD}

\author[1,2]{\fnm{Nick} \sur{Bryan-Kinns}\email{n.bryankinns@arts.ac.uk}}

\author[2]{\fnm{Bingyuan} \sur{Zhang}}
% cleozhang97@gmail.com

\author[3]{\fnm{Songyan} \sur{Zhao}}
% zhaosongyan7@gmail.com

\author[2]{\fnm{Berker} \sur{Banar}}
% b.banar@qmul.ac.uk

\affil[1]{\orgdiv{Creative Computing Institute}, \orgname{University of the Arts London}, \orgaddress{\city{London} \postcode{WC1V 7EY},  \country{UK}}}

\affil[2]{\orgdiv{School of Electronic Engineering and Computer Science}, \orgname{Queen Mary University of London}, \orgaddress{\city{London} \postcode{E1 4NS},  \country{UK}}}

\affil[3]{\orgdiv{Computer Science Department}, \orgname{Carleton College}, \orgaddress{\city{Northfield} \postcode{MN 55057},  \country{USA}}}

\abstract{
Generative AI models for music and the arts in general are increasingly complex and hard to understand. The field of eXplainable AI (XAI) seeks to make complex and opaque AI models such as neural networks more understandable to people. One approach to making generative AI models more understandable is to impose a small number of semantically meaningful attributes on generative AI models. 
\rr{This paper contributes a systematic examination of the impact that different combinations of Variational Auto-Encoder models (MeasureVAE and AdversarialVAE), configurations of latent space in the AI model (from 4 to 256 latent dimensions), and training datasets (Irish folk, Turkish folk, Classical, and pop) have on music generation performance when 2 or 4 meaningful musical attributes are imposed on the generative model. To date there have been no systematic comparisons of such models at this level of combinatorial detail. Our findings show that MeasureVAE has better reconstruction performance than AdversarialVAE which has better musical attribute independence. Results demonstrate that MeasureVAE was able to generate music across music genres with interpretable musical dimensions of control, and performs best with low complexity music such a pop and rock. We recommend that a 32 or 64 latent dimensional space is optimal for 4 regularised dimensions when using MeasureVAE to generate music across genres. Our results are the first detailed comparisons of configurations of state-of-the-art generative AI models for music and can be used to help select and configure AI models, musical features, and datasets for more understandable generation of music.}\\
\textbf{Preprint.} Springer MIR journal submission under review.\\
}

 % \note{MIR requests 5-6 keywords}



\maketitle

\section{Introduction}\label{sec1}

Music generation is a key use of AI for arts, and is arguably one of the earliest forms of AI art. However, contemporary generative music models rely increasingly on complex Machine Learning models \cite{LiteratureReviewBySturm, Herremans-2017, Carnovalini-2020} such as neural networks \cite{todd1989connectionist, eck2002first} and deep learning techniques \cite{briot2019deep, hadjeres2017deepbach, zhu2018xiaoice, huang2018music} which are difficult for people to understand and control. This makes it hard to use such models in real-world music making contexts as they are generally inaccessible to musicians or anyone besides their creator.

Making AI models more understandable to users is the focus the rapidly expanding research field of eXplainable AI (XAI) \cite{gunning_2016}. One approach to making machine learning models more understandable is to expose elements of the models to people in semantically meaningful ways. For example, using latent space regularisation \cite{pati2020attributebased} to impose semantically meaningful dimensions in latent space. To date there has been very little research on the applicability and use of XAI for the arts. Indeed, there is a lack of research on what configurations of generative AI models and datasets are more, or less, amenable to explanation.

This paper takes a first step towards understanding the link between the explanation and performance of AI models for the arts by examining what effect different AI model architectures, configurations, and training datasets have on the performance of generative AI models that have some explainable features.

\section{Related Work}

The field of eXplainable AI (XAI) \cite{gunning_2016} explores how complex and difficult to understand AI models such as neural nets can be made more understandable to people. Approaches to increasing the explainability of AI models include generating understandable explanations of AI model behaviour, structuring and labelling complex AI models to make them more understandable, and approximating the behaviour of complex models with less complex and more understandable models (ibid.).
\rr{An important element of XAI is the \emph{interpretability} of an AI model which we take to mean the ``ability to explain or to provide the meaning in understandable terms to a human'' \cite{guidotti2018survey}.}
\rr{Unfortunately the concept of explanation is ambiguous and variously defined \cite{Ciatto2020}. In Machine Learning (ML) literature XAI often refers to making the reasons behind ML decisions more comprehensible to humans.}
\rr{For example, the majority of XAI research has examined how to explain the decisions of ML classification and predictor models - see \cite{guidotti2018survey} for a thorough survey.}
\rr{There also exists a broader view of the concept of explainability in which ``explainability encompasses everything that makes ML models transparent and understandable, also including information about the data, performance, etc.'' \cite{Liao-2020} which we follow.}
\rr{In this paper we are specifically concerned with how to make AI models more interpretable for people so that they can better control the generative aspects of the AI model.}
\rr{Approaches for explaining AI models are most often tied to specific AI models and data types. There are an emerging set of approaches which are not tied to specific AI models or data types, referred to as agnostic explanators \cite{guidotti2018survey}. However, agnostic approaches, such as LIME \cite{Ribeiro2016} are concerned with building an explanation model to explain the classification/ prediction of an AI model whereas in this paper we focus on making the content of an AI model itself more interpretable so that the model can be better controlled for music generation.}

To date most XAI research has been concerned with goal-directed domains where task efficiency and transparency are key factors. For example, generating explanations for why an AI model made a medical diagnosis \cite{quellec2021explain} or how the AI models in self-driving cars work \cite{du2019look, shen2020explain}.
However, there has been little research on how XAI could be used in more creative domains such as the Arts \cite{XAIMS_NeurIps2021}. This lack of explainability typically limits the use of AI models for the Arts to the creator of the AI model and severely limits their use by artists and practitioners. Of the limited research on XAI for the Arts, \citep{gold29044} explores the presentation of visual cues between mappings in the latent space of an AI model, and \citep{McCormack2019} researches the visualisation of levels of mutual trust between an AI system and musicians in music making.
This leaves many open research questions on the use of XAI for the Arts ranging from questions about the explainability of different AI models and datasets to how to design user interfaces to navigate and manipulate explanations of generative AI models.

\rdelete{As noted in \cite{XAIMS_NeurIps2021}, the explainability of AI models for the Arts can be broken down into three categories: i) the \emph{role} of the AI from creative tool to co-creative producer of content; ii) the \emph{interaction} with the AI from non-interactive to rich forms of real time interaction; and iii) the \emph{common ground} with the AI, or what useful understanding users of an AI system might be able to form about the AI's outputs. For truly co-creative partnerships between AI and people we need to build AI models which operate as partners (role), are highly responsive (interaction), and allow people to build an in-depth understanding and intuition about what the AI is doing (common ground). Furthermore, in the Arts we need to carefully strike a balance between explaining an AI's creative output, and allowing opportunities for serendipity and surprise in the co-creation between people and AI. }

Taking music as a key form of artistic endeavour, this paper explores explainable AI for music. \rr{Musical problems addressed by AI models include composition, interpretation, improvisation, and accompaniment \cite{Pasquier2017}.} In this paper we focus on a core use of AI for music -- music composition or \emph{generation}, otherwise know as generative music. 
\rr{Music itself has a multi-level structure that ``ranges from timbre and sound through notes, chords, rhythmic patterns, harmonic patterns (e.g., cadences), melodic motifs, themes, sections, etc.'' \cite{Widmer2017}. As such, generative AI models range in purpose from generating monophonic sequences of notes (referred to as a melody), to polyphonic melodies, multivoice polyphony, accompaniment to a melody (counterpoint or chord progression), and association of a melody with a chord progression \cite{MusGenSurvey}.
However, sequencing longer term structures such as themes and sections by integrating short-term and long-term machine learning for music generation remains an open research challenge \cite{Widmer2017} which is problematic given that the ``long-term and/or hierarchical structure of the music plays an important role'' \cite{Herremans-2017}.}
\rr{Applications of generative AI range} from polyphonic classical music generation in the style of Bach \cite{hadjeres2017deepbach} to monophonic Irish Folk music generation \cite{Florian2016Algorithmic, Sturm2016MusicTM}, \rr{and include composition applications such as musical inpainting to generate a melody to fill in the musical gap between two melodies \cite{pati2019learning} and musical interpolation to generate a set of melodies which incrementally move from one melody to another \cite{roberts2018hierarchical}.} However, the complex nature of these generative models means that people often need some technical expertise and knowledge of these algorithms in order to use and adapt them effectively. This makes such approaches difficult for people, especially non-experts, to understand and manipulate. 

In this paper we focus very much on the explainability of the AI model itself and its output.
\rdelete{ rather than considering the role of the AI model or its forms of interaction.} 
In particular, we examine how semantically meaningful labels can be applied to properties of AI models to provide the opportunity for users to interpret and understand some aspects of the model and its generated output.
\rr{To date there have been few comparisons of the performance of explainable generative models for music. For example, research has compared the performance of a novel Convolutional-Variational Neural Network for music generation to other Neural Networks \cite{Koh2018} in terms of the Information Rate of generated music - a measure of musical structure. However, such comparisons compare across models, not comparing the configurations of models themselves, and do not examine a range of semantically meaningful features.}
\rr{We aim to compare the effect of meaningful labels on AI models in different configurations and with different datasets. To reduce the complexity of these combinatorial analyses we select the core music generation task of generating monophonic melodies. }
\rr{In this way we contribute the first in-depth analysis} of how different AI model architectures and datasets affect music generation when explainable attributes are used.
\rr{Future work can build on our findings to compare the effects of explainable attributes on more complex polyphonic melody generation and later accompaniment and association.} \rr{By taking this approach we improve the field's understanding of state-of-the-art deep learning generative models to help inform future generative model development and refinement - understanding where we are today informs where we might go in the future.} 

\subsection{Latent Spaces for Music Generation}\label{sec:LatentSpaces}

AI models for music generation range from probability based models such as Markov Chains \cite{Ames1989TheMP}\cite{Whorley2020} through to deep learning techniques explored in this paper \cite{pati2019learning, Kawai2020AttributesAwareDM}.
\rr{Probabilistic approaches typically offer more controllable music generation with lower computational and dataset requirements, but their outputs are often less novel than those achieved by deep learning approaches.} 
\rr{A wide variety of deep learning generative models of music have been developed in recent years \cite{MusGenSurvey}}
\rdelete{Deep learning models have become increasingly popular in generative music research}
and have been demonstrated to generate convincing musical outputs \cite{LiteratureReviewBySturm, Herremans-2017, Carnovalini-2020}. \rr{Briot et al. \cite{MusGenSurvey} provide a thorough survey of deep learning architectures and models used for music generation including Variational Auto Encoder (VAE), Restricted Boltzmann Machine (RBM), Recurrent Neural Network (RNN), Convolutional neural network (CNN), Generative Adversarial Networks (GAN), Reinforcement learning (RL), and Compound Architectures of these approaches.}
As noted by \cite{Kawai2020AttributesAwareDM}, two of the most popular deep learning approaches to generative AI are Generative Adversarial Networks (GANs) \cite{Goodfellow2014GAN} and Variational Auto-Encoders (VAEs) \cite{Kingma2013VAE}.
\rr{In this paper we examine VAEs as they}
\rdelete{latent space models such as VAE}
have been demonstrated to be capable of creative tasks including music generation \cite{yang2019deep, wang2020pianotree}, music inpainting \cite{pati2019learning}, and music interpolation \cite{roberts2018hierarchical}.
\rr{Moreover, whilst comparisons of VAE approaches have to date examined image generation in terms of computation time and re-generation accuracy \cite{Wei2020}, there has been no systematic comparison of VAEs for music generation, nor in terms of interpretable features.}
Some recent VAE systems have exposed the latent space of generative music models to users \cite{louie2020cococo, pati2019learning, Thelle2021Spire, Murray2021Latent, gold29044} as a way for users to navigate the latent space to generate music.
These approaches offer increased control of the AI models \cite{Murray2021Latent} and increased structure and labelling of the models \cite{pati2019learning}, both of which contribute to increasing explainability. Given the research interest in making latent spaces more explainable we explore what effect different AI model configurations and training datasets have on one of these approaches when explainable attributes are applied. In this paper we explore these questions for the popular VAE architecture which shows promise as a deep learning approach to music generation \cite{roberts2018hierarchical}. 

\begin{figure}
    \centering
    \includegraphics[width = 7.5cm]{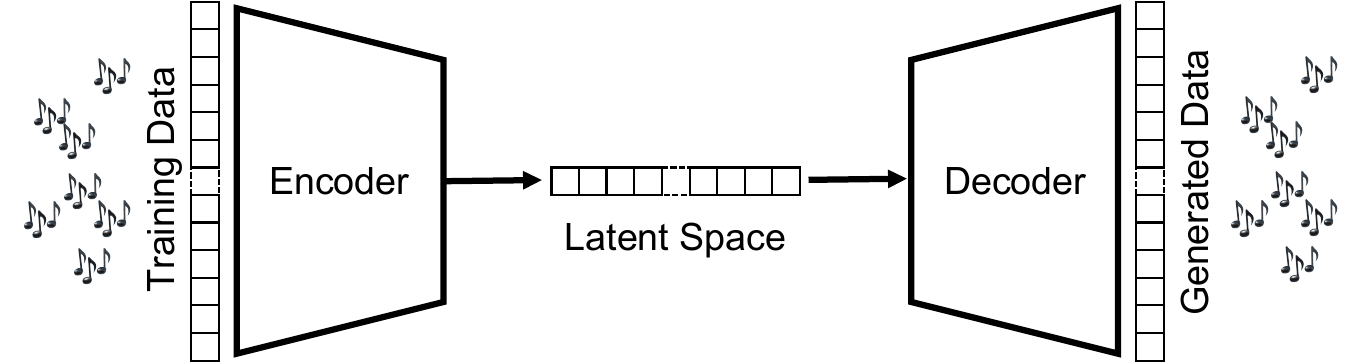}
    \caption{Variational Auto-Encoder Architecture}
    \label{fig:VAE_generic}
\end{figure}

A VAE architecture consists of i) an \emph{encoder} which encodes training data into ii) a multi-dimensional \emph{latent space} which is used by iii) a \emph{decoder} which decodes data from the latent space to generate data in the style of the training data as illustrated in Fig. \ref{fig:VAE_generic}. Modifying values of the latent space dimensions will have an effect on the generated data. The challenge for explainable VAE data generation is how to offer users meaningful control of the generated data given that the latent space is the result of unsupervised learning with no meaningful structure. \rr{There are two main approaches to attribute-based control of generative models: unsupervised disentanglement learning, and supervised regularisation methods \cite{pati2020attributebased}. However, unsupervised disentanglement necessarily requires some post-training analysis to identify the possible meaning of the disentangled dimensions (ibid.).} 

\section{Research Questions}
As outlined in previous sections, there are many approaches to music generation using deep learning models, and each year new models are added to the repertoire of music generation systems. However, to date there has been no systematic analysis of how different training datasets and AI model architectures might impact the performance of XAI models for music. For example, to date the only experiments on the effect of latent space dimensionality on model performance have been conducted on images \cite{gillette2020algorithms}.
Our core Research Question is: \emph{What effect do different AI model architectures, configurations, and training data have on the performance of generative AI models for music with explainable features}. \rr{No research has been undertaken to establish these effects to date. In answering this question we help researchers to understand the properties of state-of-the-art generative music architectures and so help to build a baseline from which to explore many more model features and generative architectures.}

This paper begins to address the core Research Question by systematically asking the following questions about the performance of VAE generative models with explainable features:
\begin{itemize}
\item[] \textbf{RQ1} What is the effect of VAE model architectures on performance?
\item[] \textbf{RQ2} What effect do the musical features imposed on the latent space have on performance?
\item[] \textbf{RQ3} What effect does the size of latent space have on performance?
\item[] \textbf{RQ4} What effect do training datasets have on performance?
\end{itemize}

\section{Methods}
Following \cite{pati2019learning} which demonstrates that VAEs are successful in generating short pieces of monophonic music we restrict our music generation to monophonic measures of music represented by 24 characters. Each character in can represent a musical note, a note continuation, or a rest.

\subsection{Candidate AI Models}\label{sec:candidateAIModels}
As a first step in understanding what effect explainable features have on the performance of generative AI model architectures we compare two representative example VAE generative music models - MeasureVAE \cite{pati2019learning} and AdversarialVAE \cite{Kawai2020AttributesAwareDM}.
Both approaches build on a VAE architecture (Section \ref{sec:LatentSpaces}) to generate music but differ in terms of how musically semantic information is applied to the music generation \rr{with MeasureVAE imposing regularised dimensions on the latent space and AdversarialVAE adding control attributes to the Decoder}. 

\subsubsection{MeasureVAE}
\begin{figure}
    \centering
    \includegraphics[width = 7.5cm]{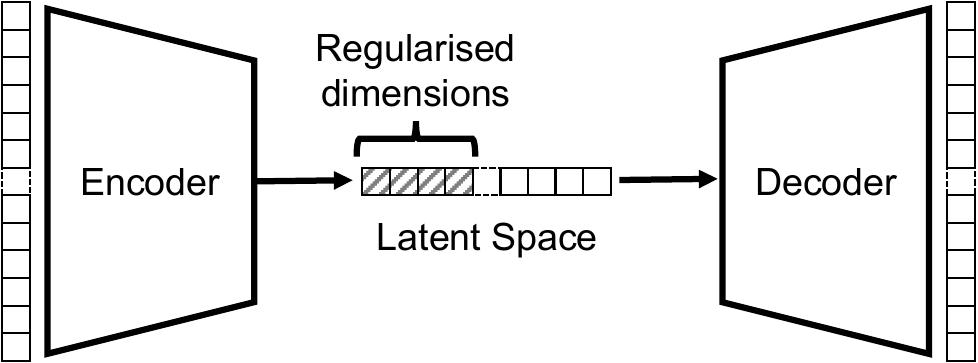}
    \caption{Variational Auto-Encoder with Latent Space Regularisation of 4 Dimensions}
    \label{fig:VAE_LSR}
\end{figure}

The popular MeasureVAE implementation\footnote{Creative Commons Attribution-NonCommercial-ShareAlike 4.0 International License.}\footnote{\url{https://github.com/ashispati/AttributeModelling}} \cite{pati2019learning, pati2} has been demonstrated to be ``successful in modeling individual measures of music'' \cite{pati2019learning}. MeasureVAE uses a bi-directional recurrent neural network (RNN) for the encoder, and a combination of two uni-directional RNNs and linear stacks for the decoder \cite{pati2019learning}.
The generated music can be varied by modifying the values of the dimensions in the latent space but the relationships between the dimensions and the music produced is not meaningful to people.
To improve the explainability of the MeasureVAE we can apply latent space regularisation (LSR) \cite{hadjeres2017glsrvae} when training the VAE. LSR has been widely used to allow more user controlled generation of images \cite{lample2018fader} and music \cite{pati2, tan2020music}. Following \citet{pati2, pati2020attributebased} we use LSR to force a small number of dimensions of the latent space to represent specific musical attributes (see Section \ref{sec:musical_features}) -- these regularised dimensions are the explainable features of the MeasureVAE model which increase the explainability of the generative model. Fig. \ref{fig:VAE_LSR} illustrates the VAE architecture with 4 regularised dimensions in the latent space. See \cite{pati2019learning} for details of the MeasureVAE Model Architecture.

\rr{In MeasureVAE, which is a typical VAE encoder-decoder architecture \cite{pati2020attributebased}, data points \textbf{x} in a high-dimensional space \textbf{X} are mapped to a low-dimensional latent space \textbf{Z} using the encoder, where latent vectors are represented with \textbf{z}. The latent vectors, \textbf{z} are mapped back to the data space, \textbf{X} via the decoder. Latent vector, \textbf{z}, is considered as a random variable and generation process is defined with the sampling processes of $\textbf{z} \sim p(\textbf{z})$ and $\textbf{x} \sim p_\theta(\textbf{x}\mid\textbf{z})$. $p_\theta(\textbf{x}\mid\textbf{z})$ is the $\theta$ parameterised decoder architecture and $p(\textbf{z})$ is the prior distribution over the latent space, \textbf{Z}, as per the variational inference. The encoder is represented with $q_\phi(\textbf{z}\mid\textbf{x})$, which is the posterior parameterised by $\phi$. In this context, the loss function is defined with the equation below as also defined in \cite{pati2020attributebased}:}

\begin{equation}
\rr{L_{VAE}(\theta, \phi) = L_{R}(\theta, \phi) + L_{KLD}(\theta, \phi)}
\label{eq:1}
\end{equation}

\rr{In \ref{eq:1}, the first term, $L_{R}$, represents the reconstruction loss, which is the L2 norm between \textbf{x}, original data vector, and \textbf{\^{x}}, its reconstruction version. The second, $L_{KLD}$, represents the KL-Divergence regularisation, typical to VAEs.}

\rr{To apply the latent space regularisation in the context of MeasureVAE, firstly an attribute distance matrix is defined, which is $D_a \in \mathbb{R}^{m \times m}$, where m is training examples in a mini-batch, as in \cite{pati2020attributebased}:}

\begin{equation}
\rr{D_a(i,j) = a(\textbf{x}_i) - a(\textbf{x}_j)}
\label{eq:2}
\end{equation}

\rr{where $\textbf{x}_i$ and $\textbf{x}_j$ represent arbitrary data vectors and $a(\cdot)$ is the calculation of any attribute for the data vector, $\textbf{x}$. Then, another distance matrix, $D_r \in \mathbb{R}^{m \times m}$ is calculated for the regularised dimension, $r$, of the latent vectors, $\textbf{z}$:}

\begin{equation}
\rr{D_r(i,j) = z_i^{r} - z_j^{r}}
\label{eq:3}
\end{equation}

\rr{where $z_i^{r}$ and $z_j^{r}$ are r-th dimension values of the arbitrary latent vectors $z_i$ and $z_j$. Lastly, the additional loss term for the latent space regularisation is defined with the following equation, as in \cite{pati2020attributebased}:}

\begin{equation}
\rr{L_{r,a} = MAE(tanh(\delta{D_r}) - sgn(D_a)}
\label{eq:4}
\end{equation}

\rr{which is added to the VAE loss in \ref{eq:1}. In \ref{eq:1}, $MAE$ is the mean absolute error, $tanh$ is the hyperbolic tangent, $sgn$ is the sign function and $\delta$ is a parameter that controls the spread of the posterior. Due to this additional term, increasing or decreasing relationships between the calculated attributes for $x_i$ and $x_j$ are similarly reflected to the relationship between $z_i^r$ and $z_j^r$, respectively. The code for this MeasureVAE implementation based on \cite{pati2020attributebased} can be found here \footnote{\url{https://github.com/bbanar2/Exploring_XAI_in_GenMus_via_LSR}}}.

\subsubsection{AdversarialVAE}
\begin{figure}
    \centering
    \includegraphics[width = 7.5cm]{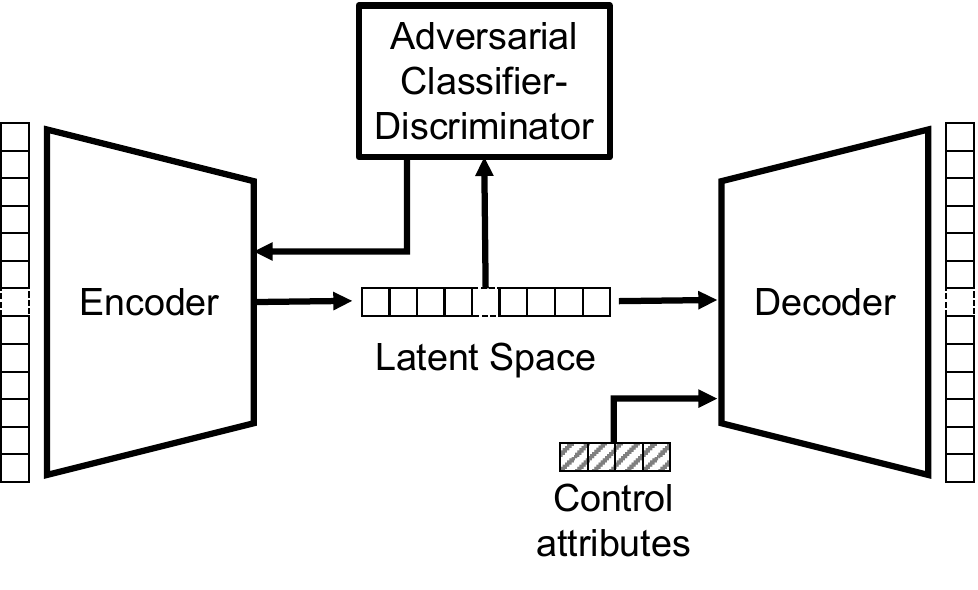}
    \caption{Variational Auto-Encoder with Adversarial Classifier and Decoder Control Attributes}
    \label{fig:VAE_Adversarial}
\end{figure}

The AdversarialVAE \cite{Kawai2020AttributesAwareDM} uses a one-layer bidirectional Gated Recurrent Unit (GRU) for the encoder \rr{followed by linear layers (MLP) for the mean and variance of the variational sampling at the latent space bottleneck}, and a two-layer GRU for the decoder \rr{followed by a linear layer (MLP)} which in contrast for MeasureVAE uses both the latent space and \rr{additional} control attributes to generate the music. The latent space itself has an adversarial classifier-discriminator added which ``induces the encoder to remove the attribute information from the latent vector'' (ibid.) as illustrated in Fig. \ref{fig:VAE_Adversarial}. In contrast to MeasureVAE where specific dimensions of the latent space are tied to semantic musical features, music generation with the AdversarialVAE is controlled by musical attributes fed to the decoder -- these are the explainable features of the AdversarialVAE model.

\rr{The AdversarialVAE model as defined in \cite{Kawai2020AttributesAwareDM}, similar to MeasureVAE, has an additional loss term on top of the original VAE loss, which is defined in equation \ref{eq:1}. The additional loss term here, denoted as $L_{D}$, is adversarial and it belongs to a separate architecture that is a classifier-discriminator which consists of linear layers with $tanh$ activations, except for the last layer where sigmoid activation is utilised as per the classification task. The objective of this additional classifier-discriminator is to determine the value of a musical attribute using discretely defined levels given the latent vector, $\textbf{z}$, of a musical sequence by learning a probability distribution $s_{\psi}$, where $\psi$ is the parameterised classifier-discriminator.}

\rr{To construct $L_{D}$, firstly $N$ many musical attributes are defined. Then, based on the training data, each attribute is quantised into $K$ many bins (specifically, $K = 8$ in this study), where $\mu$-law compression is used as in \cite{Kawai2020AttributesAwareDM} to obtain equal number of samples in each bin given the characteristics of the training data. Labels of each sample in the training data are one-hot encoded according to the quantised bin that the sample belongs to. Considering $N$ many musical metrics and $K$ many discrete levels, each sample yields in a matrix $\textbf{B} \in \mathbb{R}^{N \times K}$ for the target musical attributes, which is the output of the classifier-discriminator network. As per the adversarial objective defined in \cite{Kawai2020AttributesAwareDM}, the encoder tries to prevent the classifier-discriminator to predict to correct targets for the musical attributes, therefore the targets from the perspective of the encoder are defined as $\textbf{1} - \textbf{B}$, where $1$ is the matrix of ones as per the one-hot encoding.}

\rr{After having the $\textbf{B}$ and $\textbf{1} - \textbf{B}$ matrices, the $L_{D}$ is defined as follows:}

\begin{equation}
    \rr{L_{D}(\psi \mid \phi) = -E_{q_{\phi}(\textbf{z}\mid\textbf{x})}[log(s_{\psi}(\textbf{B}\mid\textbf{z}))] }
    \label{eq:5}
\end{equation}

\begin{equation}
    \rr{L_{D}(\phi \mid \psi) = -E_{q_{\phi}(\textbf{z}\mid\textbf{x})}[log(s_{\psi}(\textbf{1} - \textbf{B}\mid\textbf{z}))]}
    \label{eq:6}
\end{equation}

\rr{where $\phi$ is the parameterised encoder architecture, $\psi$ is the parameterised classifier-discriminator, $q_{\phi}(\textbf{z}\mid\textbf{x})$ is the posterior distribution denoting the encoder in accordance with the notation in equation \ref{eq:1}.}

\rr{Then, the overall loss becomes:}
\begin{equation}
    \rr{L(\phi,\theta\mid\psi) = L_{R}(\theta, \phi) + L_{KLD}(\theta, \phi) + L_{D}(\phi\mid\psi)}
    \label{eq:7}
\end{equation}

\begin{equation}
    \rr{L(\psi\mid\phi) = L_{D}(\phi\mid\psi)}
    \label{eq:7}
\end{equation}

\rr{following a similar notation as above.}

\rr{See \cite{Kawai2020AttributesAwareDM} for details of the AdversarialVAE Model Architecture and the repository here\footnote{\url{https://github.com/RadixBupleuri/VAEs}} for the implementation of AdversarialVAE based on \cite{Kawai2020AttributesAwareDM}}.

\subsection{Datasets}\label{sec:MusicDatasets}
Generative AI music models are typically trained and evaluated on one musical dataset such as monophonic Irish folk melodies \cite{Sturm2016MusicTM} which have been used to train and test MeasureVAE \cite{pati2019learning}. However, as noted in \cite{Banar2022GPT2}, different musical genres have different musical features which may have an impact on the performance of a generative AI model and potentially its explainability. 

In this paper we use the frequently used Irish Folk dataset \cite{Sturm2016MusicTM} and compare and contrast this with datasets of Turkish folk music, pop music, and classical music as described in this section. Table \ref{tab:datasets} presents key features of the datasets used including their musical features from Section \ref{sec:musical_features}.

\subsubsection{Irish Folk dataset}\label{sec:IrishFolkDataset}
The Irish Folk dataset contains 20,000 monophonic Irish folk melodies \cite{Sturm2016MusicTM}\footnote{\url{https://github.com/IraKorshunova/folk-rnn}} from which 5.6m notes are extracted for these experiments. The dataset has the highest note range and density of the music used in these experiments meaning that it is the most complex musically. It is also by far the largest dataset used in this experiment and is commonly used in generative AI research.

\subsubsection{Turkish Makam dataset}
The Turkish Makam dataset \cite{Karaosmanoglu2012}\footnote{\url{https://github.com/MTG/SymbTr/releases/tag/v2.0.0}} as used in \cite{Dzhambazov2016OnTU} which consists of approximately 2,200 musical scores related to Turkish makam music which is a form of Turkish folk music. This results in approximately 755k musical notes of monophonic folk songs. The Turkish Makam dataset has similarly high mean note density, note range, and rhythmic complexity to the Irish Folk music dataset suggesting similarly high musical complexity. The Turkish Makam dataset is the smallest dataset using in these experiments.

\subsubsection{Muse Bach dataset}
MuseData\footnote{\url{https://musedata.org}} consists of Baroque to early classical music, including both monophonic and polyphonic instrumental pieces. Given the wide range of styles contained in MuseData we selected all pieces composed by Bach in MuseData to provide a coherent style of music given. 593 Bach pieces were extracted from MuseData resulting in 4,531 single-lined melodies and almost 1m musical notes which we refer to as the Muse Bach dataset. The Muse Bach dataset has the lowest mean average interval jump of the datasets used, and middling note density, range, and complexity.

\subsubsection{Lakh Clean dataset}
The Lakh dataset \cite{Raffel2016}\footnote{\url{ https://colinraffel.com/projects/lmd/}} contains 176,581 unique MIDI files. For this experiment we use a subset of the Lakh dataset - the Clean MIDI (sub)dataset which contains pieces by 2,199 artists.
The distribution of the genres in the Clean MIDI dataset is: 33\% Pop, 32\% Rock, 13\% Jazz and Blues, 10\% R\&B, and 12\% other, providing a dataset of contemporary popular music. Almost 7k monophonic clips were extracted from these pieces resulting in approximately 1.7m notes which we refer to as the Lakh Clean dataset. This dataset has the lowest mean note range and rhythmic complexity of the datasets using in these experiments, suggesting that it contains some of the least musically complex music.

\subsubsection{Data Preparation}
Each dataset was converted into a measure based ABC format using the midi2ABC functions in EasyABC\footnote{\url{https://github.com/jwdj/EasyABC/}}. Each measure is represented by 24 characters including notes names, and continuation and rest tokens. As the VAE models in this experiment work with monophonic melodies, single line melodies were extracted from the datasest using EasyABC. All musical instruments were then separated into separate files and any remaining chords were converted into single notes based on the chord's highest pitch.

\begin{sidewaystable}
\sidewaystablefn%
\begin{center}
\begin{minipage}{\textwidth}
  \caption{Summary statistics of the datasets. \label{tab:datasets}}
\begin{tabular*}{\textwidth}{@{\extracolsep{\fill}}cccccccccccc@{\extracolsep{\fill}}}

\toprule
{Dataset}    & 
\multicolumn{2}{c}{{Note Density}} & 
\multicolumn{2}{c}{{Note Range}} &
\multicolumn{2}{c}{{Rhy. Complexity}} & 
\multicolumn{2}{c}{{Avg. Int. Jump}} &
{Genres} &
{No. of Notes}\\
 
 & Mean & STD & Mean & STD & Mean & STD & Mean & STD & & \\
\midrule

{Muse Bach} & 3.206 & 3.346 &3.336  &4.073  &1.742 &2.150&0.761  &1.933 & Classical  & 965,244\\
 
{Lakh Clean} & 2.504 &3.178 &1.515 &3.381 &1.462 &2.146 &2.865  &2.840 & Mainly Pop & 1,697,053\\
{Turkish Makam} &6.609 & 3.161 &4.972  &2.828  &3.907 & 2.456  &1.564  &0.818 & Folk & 755,785\\
{Irish Folk} &6.765  & 2.056 &7.809  &3.440   &3.756 &1.998  &2.653  &2.653 & Folk & 5,662,498\\
\botrule

\end{tabular*}

\end{minipage}
\end{center}
\end{sidewaystable}

\subsection{Musical Features}\label{sec:musical_features}

There are many musical features that could be imposed on music generation. For example, the popular jSymbolic \cite{Cory2006} offers analysis of 246 unique musical features. In this research we use a subset of the most frequently used features in music research, and follow \cite{XAIMS_NeurIps2021} to select the following musical attributes:
\begin{itemize}
\item \textbf{Note Density (ND)} -- the number of notes in a measure;
\item \textbf{Note Range (NR)} -- the highest minus lowest pitch in a measure;
\item \textbf{Rhythmic Complexity (RC)} -- how syncopated a musical measure is \cite{Toussaint02amathematical};
\item \textbf{Average Interval Jump (AIJ)} -- the average of the absolute difference between adjacent notes in a measure.
\end{itemize}

These features cover both rhythmical properties (ND and RC) and melodic properties (NR and AIJ). They are used to i) characterise the musical properties of the datasets (Section \ref{sec:MusicDatasets}) used to train the AI models; and ii) as attributes of control of music generation - because the features have some musical meaning they serve to increase the explainability of the AI models.

\section{Experimental Setting: Comparing VAE Model Architectures}\label{sec:compareVAEmodels}
MeasureVAE and AdversarialVAE were compared using the Irish Folk dataset for training to explore RQ1. Of the 20,000 monophonic Irish folk melodies, 14,000 were used as the training set, 3,000 as test sets, and 3,000 as validation sets. Models were compared in terms of the evaluation metrics outlined in Section \ref{sec:VAEEvaluationMetrics}.

The experiment learning rate was set to 1e-4 (optimized using Adam\cite{kingma2014adam}). Both models were trained on a GPU for a total of 50 iterations with a batch size of 64 for all data. $\alpha=0.1$, $\beta=0.1$, $\gamma= 0.2$ was used in the VAE loss function. As MeasureVAE uses 256 dimensions of latent space whereas AdversarialVAE uses 128 dimensions testing was undertaken with both 128 and 256 dimensions \cite{gillette2020algorithms}. Both models had 4 musical features imposed on them (from Section \ref{sec:musical_features}).

\rr{We use musical measures for generative output and training in keeping with state-of-the-art music generation research \cite{pati2019learning, pati2020attributebased} and typical of the musical elements used in current generative AI tasks. Each measure is represented by 24 characters which include note names such as A3, G5, and so on, continuation tokens, and rest tokens.}

\subsection{Evaluation Metrics}\label{sec:VAEEvaluationMetrics}
We evaluate the AI models in terms of standard measures of: 

\begin{itemize}
\item \textbf{Reconstruction Accuracy} -- how well the model can reconstruct any given input.
We aim to maximise this. This is calculated by comparing the difference between the input melody and generated melody, and averaging this over the whole dataset. \rr{Reconstruction accuracy is defined as follows:}

\begin{equation}
    \rr{RA(\textbf{x},\textbf{\^{x}}) = \frac{100}{N}\sum_{i=1}^{N} \frac{1}{M_i}\sum_{j=1}^{M_i}Check(x_{ij}, \hat{x}_{ij})}
    \label{eq:9}
\end{equation}

where $x_i$ is the input sequence, $\hat{x}_{i}$ is the reconstructed sequence, N is the number of samples in a dataset and M is the sequence length. The Check function compares two corresponding elements in the original and the reconstructed sequences as follows:

\begin{equation}
    \rr{Check(x_{ij}, \hat{x}_{ij}) = \begin{cases} 
      1, & x_{ij} = \hat{x}_{ij} \\
      0, & o.w.
   \end{cases}}
    \label{eq:10}
\end{equation}

\item \textbf{Reconstruction Efficiency} -- how well the model generates music \rr{with respect to the characteristics of its training dataset and also the provided input sequence} when musical parameters are changed. We aim to maximise this. \rr{To calculate this measure, we split our data into two categories, where an attribute $a_r\ge0$ and $a_r<0$. Then, we calculate the mean latent vectors $\textbf{z}_\textbf{{a}}$ and $\textbf{z}_\textbf{a0}$ for each of these subsets, respectively. This procedures provides us with a general picture of latent vectors with respect to the presence of the musical attribute. Then, using these mean vectors, for each sample in our data, we get their latent vectors, $\textbf{z}$ and apply the following interpolation:}

\begin{equation}
    \rr{\textbf{z}_\textbf{{resulting}} = \textbf{z} + \mu(\textbf{z}_\textbf{{a}} - \textbf{z}_\textbf{a0})}
\end{equation}

\rr{where the $\mu \in \{-0.5, -0.4, \dots, 0.4, 0.5\}$ with 11 possible values. Then, $\textbf{z}_\textbf{{resulting}}$ vectors are decoded into generated music sequences and for each generated music sequence, $\textbf{\^{x}}$, and input music sequence, $\textbf{x}$, we check the cosine similarity between these sequences using the following formula:}

\begin{equation}
    \rr{CS(\textbf{x}, \textbf{\^{x}}) = \frac{\textbf{x} \cdot \textbf{\^{x}}}{\| \textbf{x} \| \| \textbf{\^{x}} \|}}
\end{equation}

\rdelete{This is calculated by interpolating across the models from attribute minimum of $u= -0.5$ to maximum $+0.5$ and calculating the cosine similarity between the generated melody and the training melody.}
Each musical attribute is interpolated separately, and then the average similarity of each interpolation is calculated.

\item \textbf{Attribute Independence} -- how resilient an attribute is to change by other attributes. We aim to maximise this. This is calculated as \rr{getting the maximum value for Spearman's correlation coefficient between the attribute value, $a(\textbf{x})$, and each dimension of the latent space, $z_d$, \cite{myers2004spearman}. Then, these correlation coefficients are averaged for all of the musical attributes. } 
\rdelete{the difference in attribute values of the musical measure before and after the attribute is changed for all vectors in the latent space.}

\end{itemize}

\rr{Implementations of the Reconstruction Accuracy and Attribute Independence are included in these repositories} \footnote{\url{https://github.com/bbanar2/Exploring_XAI_in_GenMus_via_LSR}} \footnote{\url{https://github.com/RadixBupleuri/VAEs}} \rr{as in \cite{pati2020attributebased} and \cite{Kawai2020AttributesAwareDM}}.

\subsection{Results}

\subsubsection{Reconstruction Accuracy}

Table \ref{table:MA_ReconstructionAccuracy} shows the Reconstruction Accuracy scores for 128 and 256 latent space dimensions for MeasureVAE and AdversarialVAE. Results show that MeasureVAE reconstruction's accuracy outperformed the AdversarialVAE in both 128 and 256 dimension configurations, achieving a high of 99.6\% for the validation set with 256 dimensions. Both models performed better in 256 dimensions than in 128 dimensions. This may be because the higher number of dimensions makes it easier to decompress the latent space.

\begin{table}[h]
\begin{center}
\begin{minipage}{174pt}
\caption{The Reconstruction Accuracy of MeasureVAE and AdversarialVAE models on training, test and validation data from Irish Folk dataset.}
\label{table:MA_ReconstructionAccuracy}
\resizebox{\textwidth}{!}{
\begin{tabular}{@{}cccc@{}}
\toprule
Dims. & Dataset & \multicolumn{2}{c}{Reconstruction Accuracy (\%)}\\
 & & MeasureVAE & AdversarialVAE\\
\midrule
    & Train   & 97.359  & 94.721  \\
128    & Test   & 96.743  & 93.626  \\
    & Validate  & 96.207  & 93.572  \\
\midrule
    & Train   & 99.824  & 95.866  \\
256    & Test   & 99.459  & 95.030  \\
    & Validate  & 99.674  & 95.268  \\
\botrule
\end{tabular}
}
\end{minipage}
\end{center}
\end{table}

\subsubsection{Reconstruction Efficiency}

\begin{figure}
    \centering
    \includegraphics[width = 7.5cm]{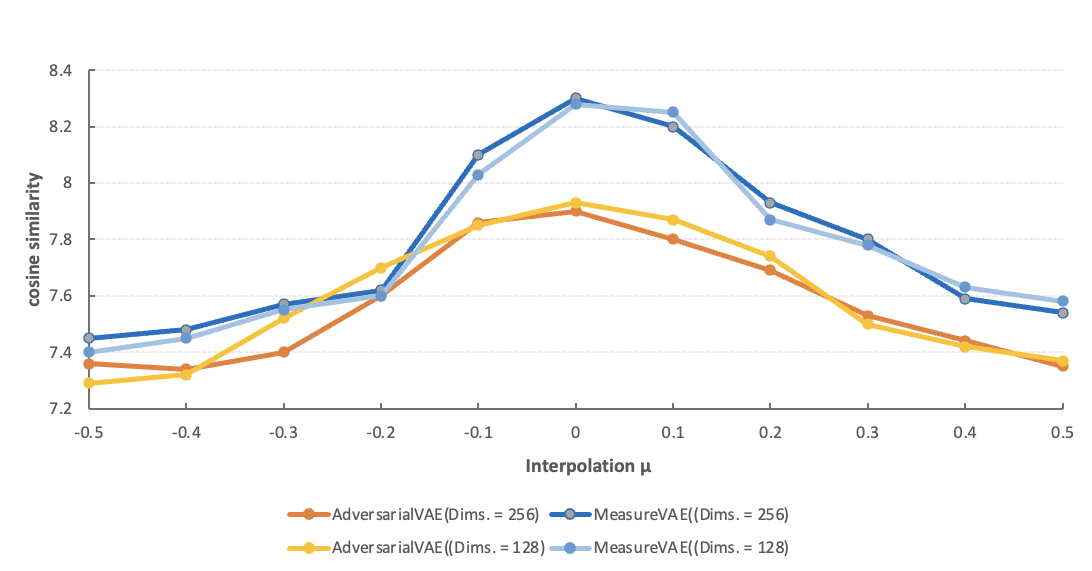}
    \caption{Reconstruction Efficiency for MeasureVAE and AdversarialVAE with 128 and 256 latent dimensions.}
    \label{fig:ReconstructionEfficiency}
\end{figure}

Fig. \ref{fig:ReconstructionEfficiency} illustrates the comparative Reconstruction Efficiency for MeasureVAE and AdversarialVAE with 128 and 256 latent dimensions, as summarised in Table \ref{table:MA_MeanStandardDeviation}. Results show that the number of dimensions of latent space (128 or 256) did not have a noticeable effect on the Reconstruction Efficiency. Regardless of the number of dimensions, MeasureVAE had higher Reconstruction Efficiency than AdversarialVAE, with largest difference between Reconstruction Efficiency at $\mu=0.0$.

\begin{figure*}[htbp]
\centering
\subfigure[Original Input]
{
\includegraphics[width=3.6cm]{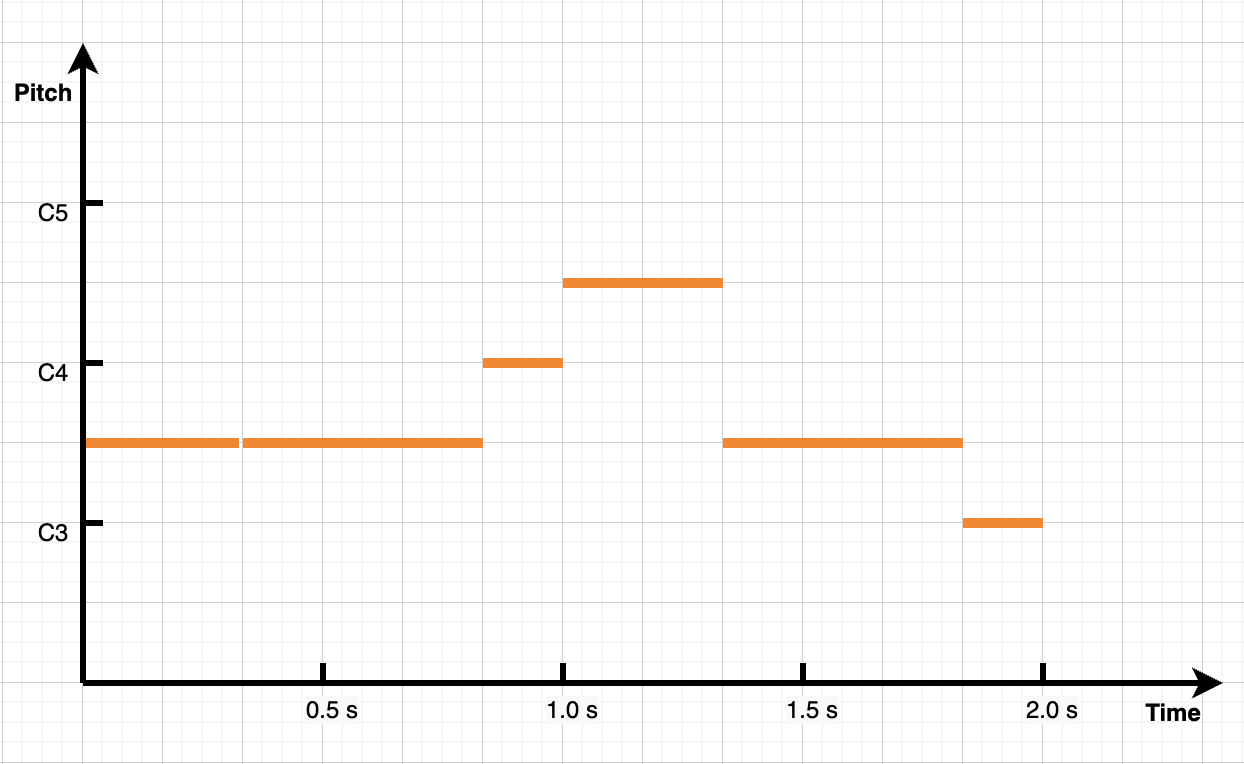}
}
\\
\subfigure[Increase NR]
{
\includegraphics[width=3.6cm]{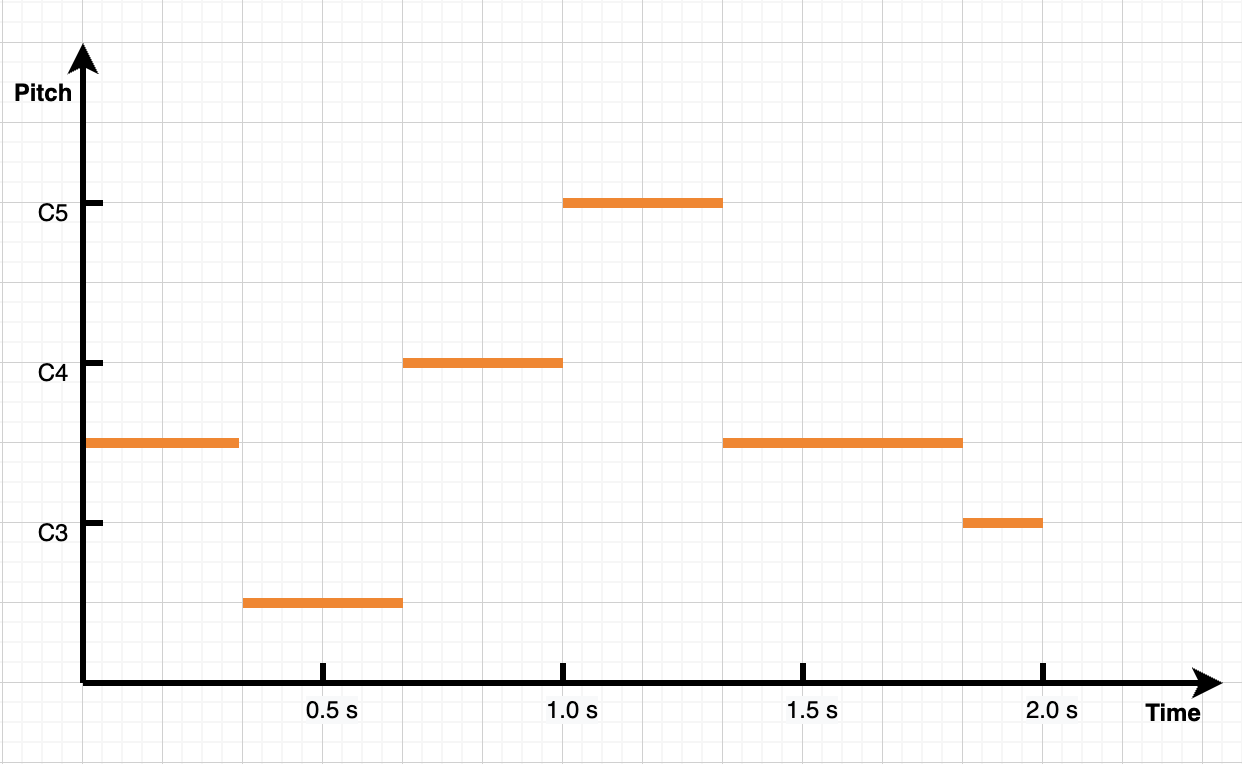}
}
\subfigure[Increase ND]
{
\includegraphics[width=3.6cm]{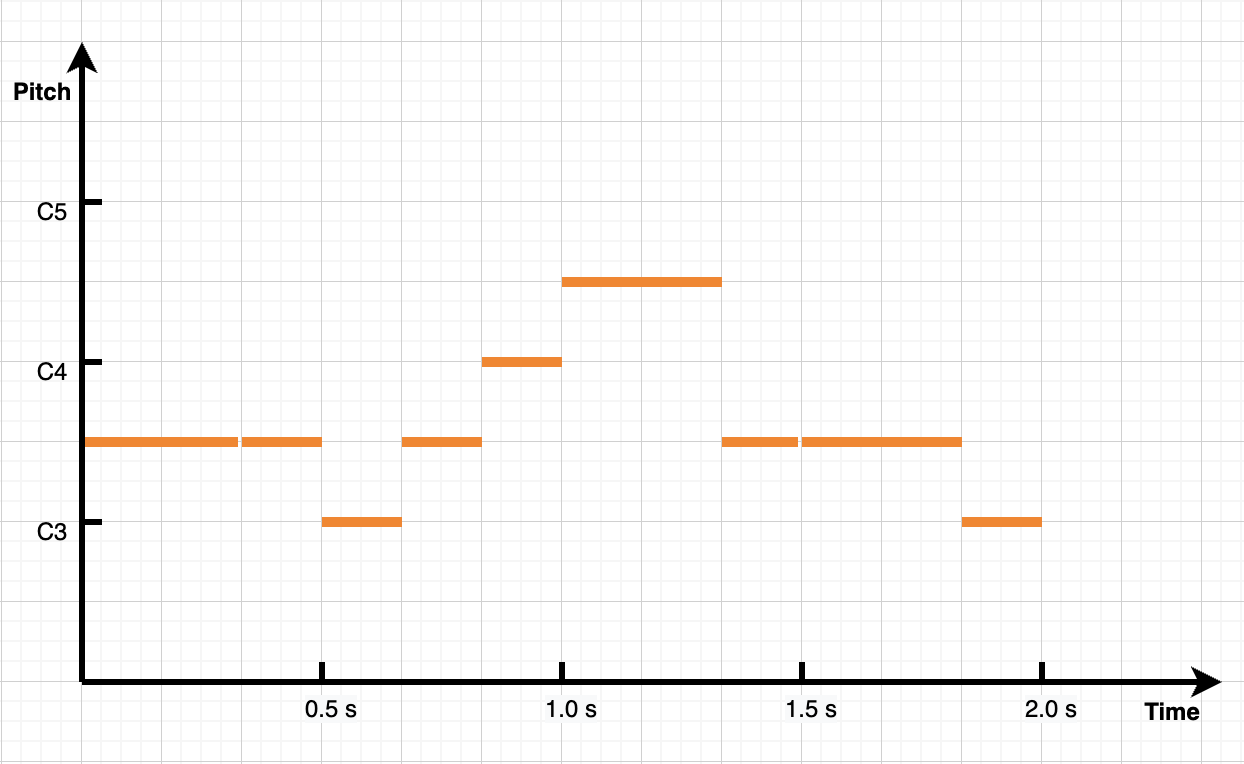}
}
\subfigure[Increase RC]
{
\includegraphics[width=3.6cm]{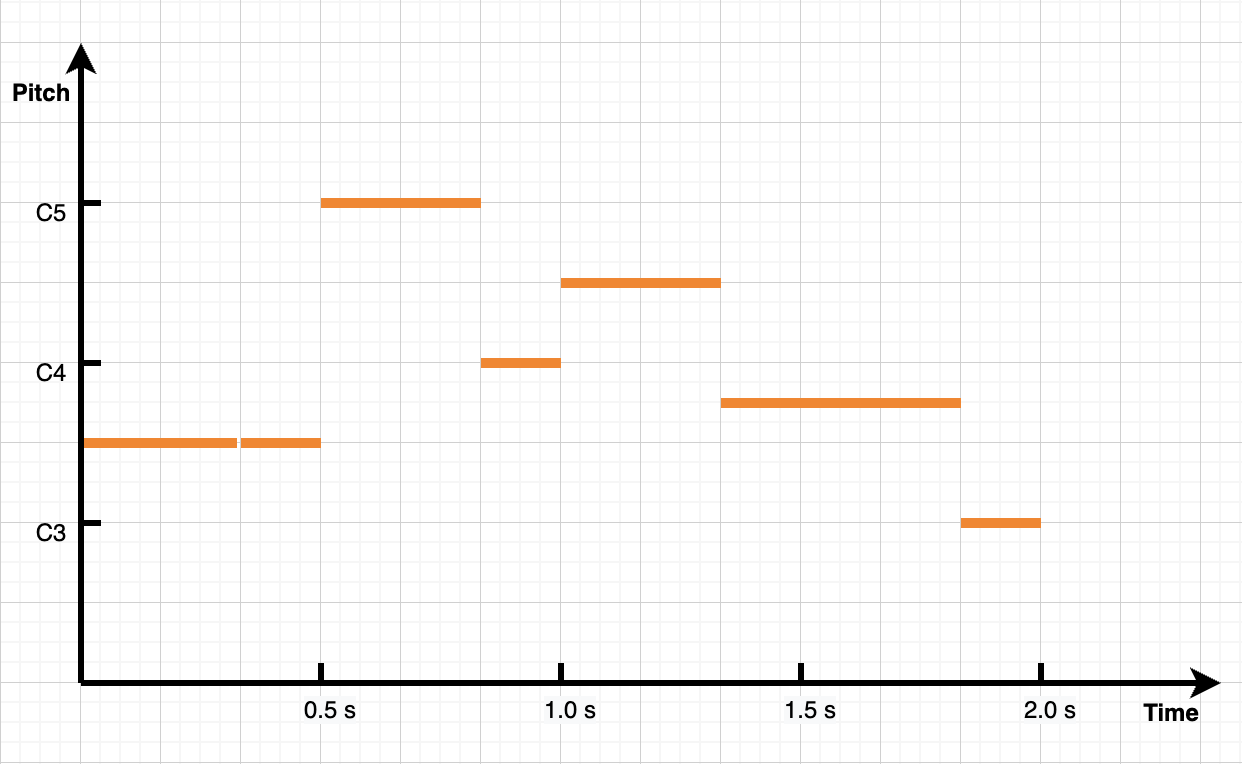}
}
\subfigure[Increase AIJ]
{
\includegraphics[width=3.6cm]{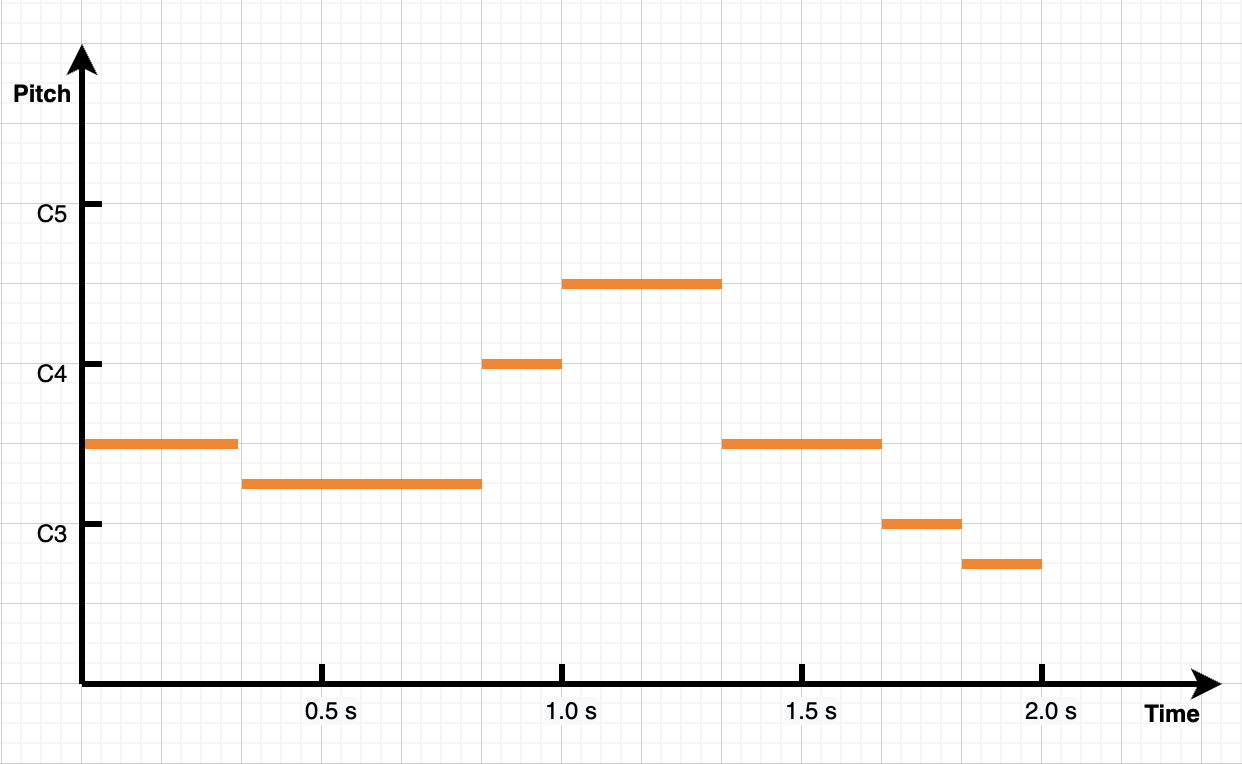}
}
\caption{Music generated by MeasureVAE for input (a): (b) to (e) show variation of four musical attributes for $\mu= +0.3$. }
\label{fig:midiM}
\end{figure*}

\begin{figure*}[htbp]
\centering
\subfigure[Original Input]
{
\includegraphics[width=3.6cm]{images/music_picture/music_origin.png}
}
\\
\subfigure[Increase NR]
{
\includegraphics[width=3.6cm]{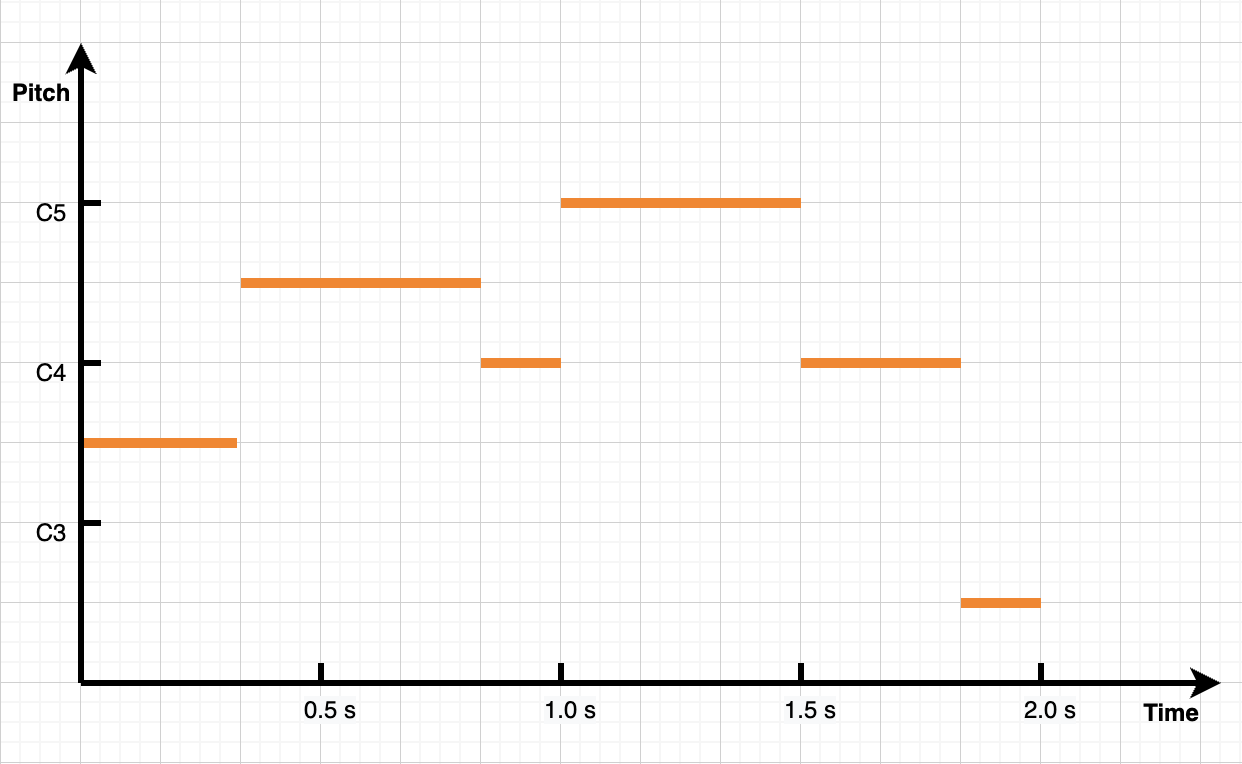}
}
\subfigure[Increase ND]
{
\includegraphics[width=3.6cm]{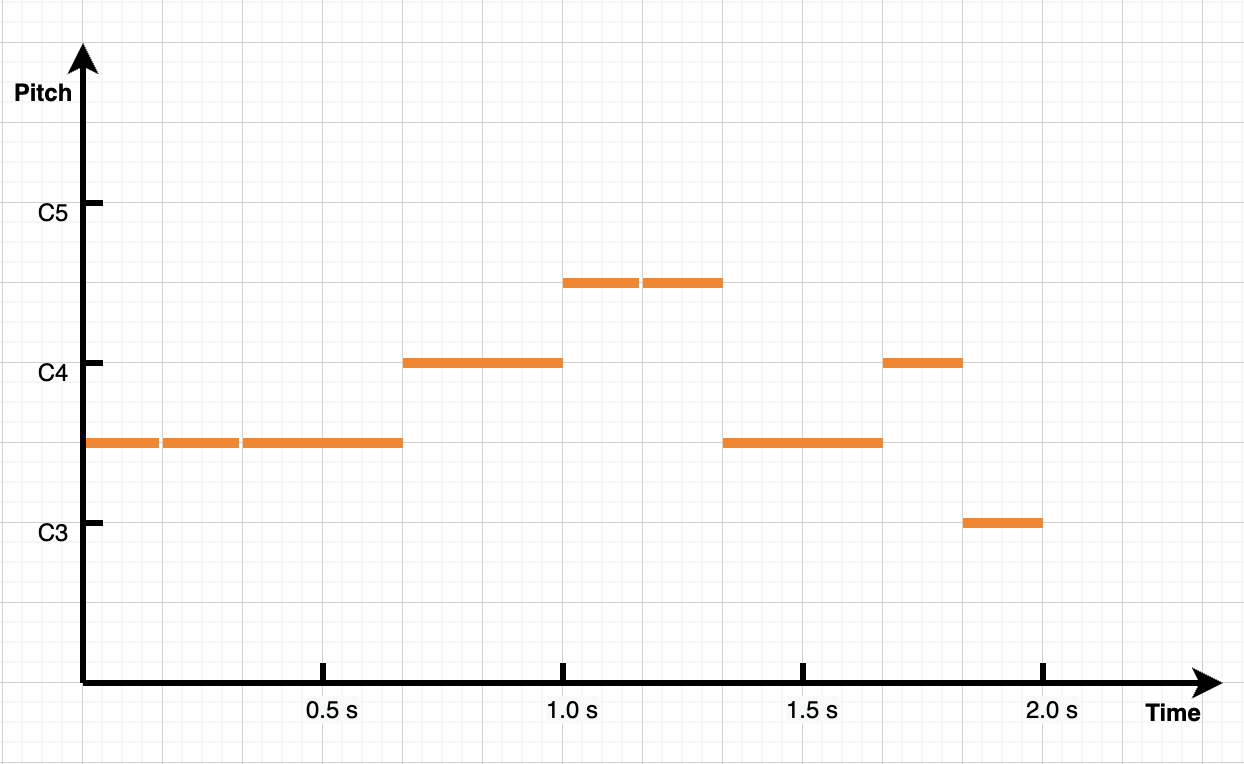}
}
\subfigure[Increase RC]
{
\includegraphics[width=3.6cm]{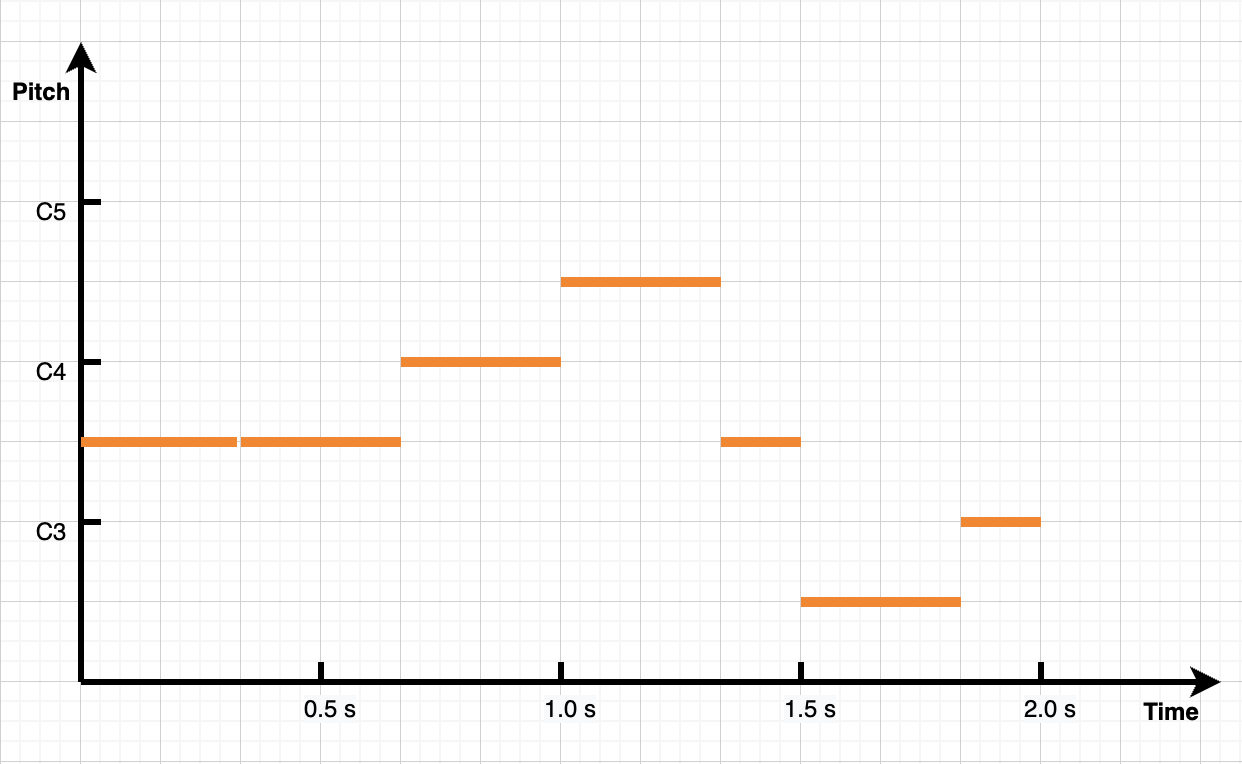}
}
\subfigure[Increase AIJ]
{
\includegraphics[width=3.6cm]{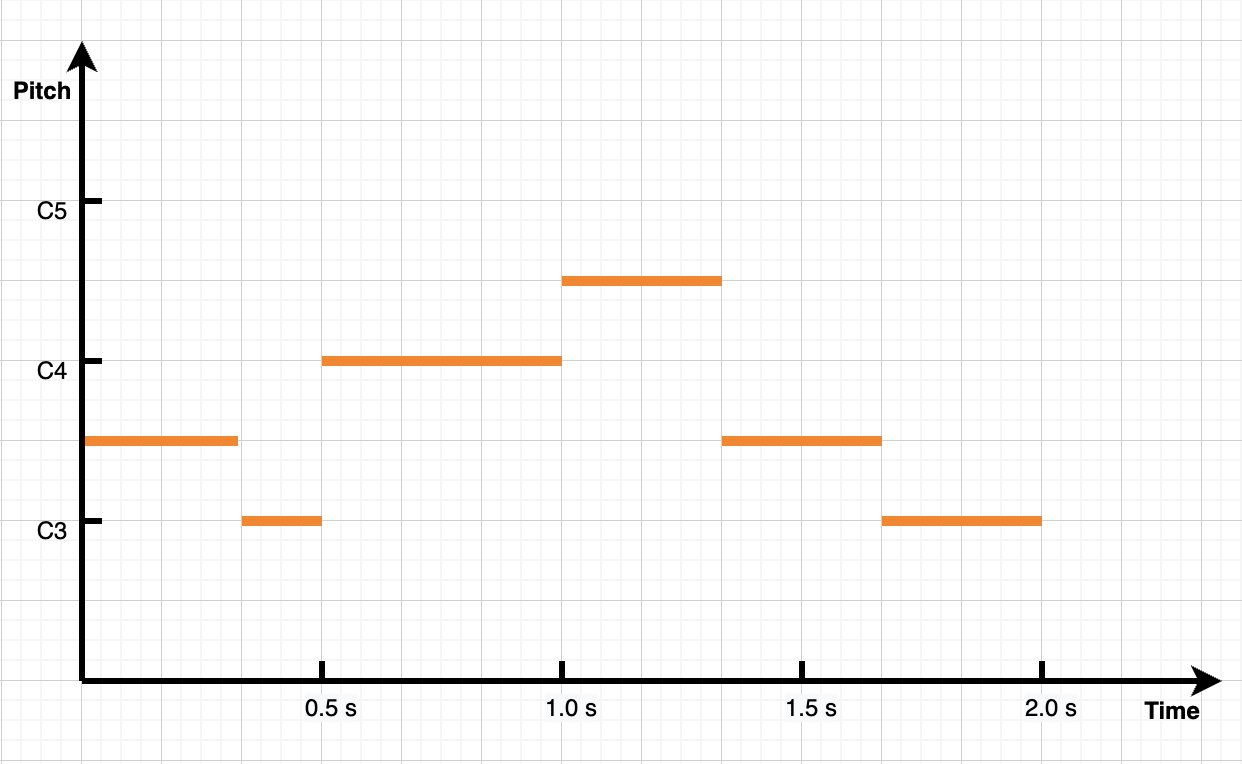}
}

\caption{Music generated by AdversarialVAE for input (a): (b) to (e) show variation of four musical attributes for $\mu= +0.3$. }
\label{fig:midiA}
\end{figure*}

\textbf{Generated Outputs.}
Figs. \ref{fig:midiM} and \ref{fig:midiA} illustrate example outputs of the MeasureVAE and AdversarialVAE models respectively for 256 latent dimensions. For both Figs. (a) shows the input notes for the AI Model, and (b) to (e) show the the melody produced by the model after interpolating for one musical attribute at $\mu=+0.3$. Each shows a the melody generated for a different musical feature: (b) Note Range increase; (c) Note Density increase; (d) Rhythmic Complexity increase; (e) Average Interval Jump increase.

Inspecting Figs. \ref{fig:midiM} and \ref{fig:midiA} suggests that the two models produce different sounding music to each other for increased Note Range (b) with MeasureVAE producing a measure with larger changes between the notes. Furthermore, increasing the Note Range MeasureVAE also increased the Average Interval Jump in Fig. \ref{fig:midiM}b, whereas AdversarialVAE produced a measure which shifted most of the original melody upwards in pitch expect for the final note which was shifted down to produce the required increase in Note Range in Fig. \ref{fig:midiA}b. This difference is illustrated by calculating the Spearman's correlation $r$ between the musical attribute values. In this case we see that the correlation between increase in Note Range and the generated measure's AIJ is $r = 0.286$ for MeasureVAE (weak correlation) and $r = 0.154$ for AdversarialVAE (no correlation) i.e. AIJ increases weakly with increases in NR for MeasureVAE but not for AdversarialVAE.

For Note Density (c), MeasureVAE produces music with a higher Note Density than AdversarialVAE when there is an increase in Note Density attribute applied to the generation. Interestingly MeasureVAE achieves the increased Note Density by adding an upward run of notes to the measure which also increases the Rhythmic Complexity whereas AdversarialVAE's increase in Note Density reduces Rhythmic Complexity compared to the original. Calculating Spearman's correlation again we see that in this case of increasing Note Density, the correlation of increased ND to output RC for MeasureVAE is $r = 0.341$ (weak correlation) and AdversarialVAE $r = 0.178$ (no correlation). This suggests that RC increases weakly when ND is increased with MeasureVAE, but not with AdversarialVAE.

In contrast to NR and ND, both AI models generate similar music to each other for increased Rhythmic Complexity (d) and also for increased Average Interval Jump (e).

\begin{table}[h]
\begin{center}
\begin{minipage}{174pt}
\caption{The Mean and Standard Deviation of Reconstruction Efficiency for MeasureVAE and AdversarialVAE models with 128 and 256 latent dimensions.}
\label{table:MA_MeanStandardDeviation}
\resizebox{\textwidth}{!}{
\begin{tabular}{@{}cccc@{}}
\toprule
Dims.  & Evaluation & \multicolumn{2}{c}{Reconstruction Efficiency}\\
 & & MeasureVAE & AdversarialVAE\\
\midrule
\multirow{2}*{128}
    & Mean   & 7.760  & 7.589  \\
    & S.D.   & 0.307  & 0.234  \\
\midrule
\multirow{2}*{256}
    & Mean   & 7.775  & 7.567  \\
    & S.D.   & 0.306  & 0.213  \\
\botrule
\end{tabular}
}
\end{minipage}
\end{center}
\end{table}

\subsubsection{Attribute Independence}

Table \ref{table:MA_AttributeIndependence} shows the results for Attribute Independence tests of MeasureVAE and AdversarialVAE for musical attributes Note Range (NR), Note Density (ND), Rhythmic Complexity (RC) and Average Interval Jump (AIJ). Results indicate that AdversarialVAE performs better than MeasureVAE for Attribute Independence for all attributes and latent dimensions except for Note Range with 128 latent dimensions where MeasureVAE performs marginally better. Note Density had the highest Attribute Independence for both models and for both 128 and 256 dimensions. This may be because ND is a measure of the number of notes in a measure meaning that it is easier to distinguish compared to other metrics such as Rhythmic Complexity which relies on the ability to differentiate between different musical beat types.
Rhythmic Complexity showed the largest difference between MeasureVAE (0.878) and AdversarialVAE (0.943) for 128 dimensions, whereas Average Interval Jump shows the largest difference between MeasureVAE (0.765) and AdversarialVAE (0.875) for 256 dimensions. The higher independence of AdversarialVAE attributes may be due to the use of the adversarial classifier-discriminator to impose musical attributes and the additional phase in the training process that optimises the decisions rather than trying to lower the loss function's value. 

\begin{table}[h]
\begin{center}
\begin{minipage}{174pt}
\caption{The Attribute Independence of MeasureVAE and AdversarialVAE models for Note Range, Note Density, Rhythmic Complexity and Average Interval Jump Attributes.}
\label{table:MA_AttributeIndependence}
\resizebox{\textwidth}{!}{
\begin{tabular}{@{}cccc@{}}
\toprule
Dims. & Attributes & \multicolumn{2}{c}{Attribute Independence}\\
 & & MeasureVAE & AdversarialVAE\\
\midrule
\multirow{4}*{128}
    & NR   & 0.945  & 0.941  \\
    & ND  & 0.976  & 0.986  \\
    & RC   & 0.878  & 0.943  \\
    & AIJ  & 0.870  & 0.899  \\ 
\midrule
\multirow{4}*{256}
    & NR   & 0.928  & 0.936  \\
    & ND  & 0.969  & 0.981  \\
    & RC   & 0.812  & 0.938  \\
    & AIJ  & 0.765  &  0.875  \\
\botrule
\end{tabular}
}
\end{minipage}
\end{center}
\end{table}

\section{Experimental Setting: Latent Space Configuration and Training Datasets}

In this work we are interested in contributing towards understanding how generative models which create music in given styles can be better interpreted and manipulated by users. To this end we now explore the performance of MeasureVAE in more detail as it has higher Reconstruction Accuracy and Reconstruction Efficiency than AdversarialVAE (Section \ref{sec:compareVAEmodels}). 

In this section we examine the impact that different different configurations of musical (explainable) features (RQ2), sizes of latent spaces (RQ3), and different training datasets (RQ4) might have on the performance and explainability of MeasureVAE. To examine this systematically we undertook a combinatorial experiment examining the effect of musical dataset (n=4), number of latent dimensions (n=7), and number of regularized musical attributes (n=2) on evaluation metrics (Section \ref{sec:ConfigurationEvaluationMetrics}):

\begin{itemize}
\item \textbf{Datasets} - Muse Bach, Lakh Clean, Turkish Makam, Irish Folk datasets (Section \ref{sec:MusicDatasets}) - to compare a range of musical genres;
\item \textbf{Latent dimensions} - 4, 8, 16, 32, 64, 128, and 256 - to capture a typical range of latent space sizes;
\item \textbf{Regularised dimensions} - 2 or 4 - using musical features (Section \ref{sec:musical_features}) in the latent space - ND\&RC, NR\&AIJ, or ND\&NR\&RC\&AIJ.
\end{itemize}

For each combination of the above we trained a MeasureVAE model for 25 epochs. We use Adam \cite{kingma2014adam} as the optimizer of the model with learning rate = 1e-5, $\beta_1$ = 0.9999 and $\epsilon$ = 1e-8. The model is trained on a single rtx6000 GPU following a similar setting of \cite{pati2}, taking on average of 2.5 hours per epoch.

\subsection{Evaluation Metrics}\label{sec:ConfigurationEvaluationMetrics}
We evaluate the combinations of datasets, latent space dimensions, and regularised dimensions in terms of standard measures of:
\begin{itemize}
\item \textbf{Reconstruction Accuracy} -- how well the model can reconstruct a given input - as in Section \rr{\ref{sec:VAEEvaluationMetrics}}.

\item \textbf{Loss} -- loss scores are calculated by the sum of VAE loss (KL-divergence and reconstruction loss, typical to VAE architectures \cite{VAEoriginal}) and the loss of the latent space regularization \cite{pati2, pati2020attributebased, XAIMS_NeurIps2021} \rr{ - as in Section \ref{sec:candidateAIModels}}. We aim to minimize this.
\item \textbf{Attribute Interpretability} -- how well a musical attribute can be predicted using only one LSR dimension in the latent space \cite{pati2, pati2020attributebased} - we aim to maximise this. We suggest that higher Interpretability scores contribute to better explainability it indicates less entangled semantic dimensions cf. \cite{adel_interpretability}.

\end{itemize}

\subsection{Results}

Tables \ref{table:MeasureVAELossReconstruction} and \ref{table:MeasureVAEInterpretability} show the results for the combinatorial experiment including datasets (Muse Bach, Lakh Clean, Turkish Makam, Irish Folk), latent dimensions (4, 8, 16, 32, 64, 128, 256), regularised dimensions (ND, NR, RC, AIJ), Loss Scores and Reconstruction Accuracy Scores (Tables \ref{table:MeasureVAELossReconstruction}), and musical attribute Interpretability scores (Table \ref{table:MeasureVAEInterpretability}).

\subsubsection{Reconstruction Accuracy and Loss}

Table \ref{table:MeasureVAELossReconstruction} shows that with 32 or more latent space dimensions MeasureVAE achieves Reconstruction Accuracy scores above 99\% and Loss scores below 0.2 for all datasets and number of regularised dimensions. This \emph{suggests that MeasureVAE is capable of generating music across folk, pop, rock, jazz and blues, R\&B, and classical music}. 

\begin{figure}
    \centering
    \includegraphics[width = 7.5cm]{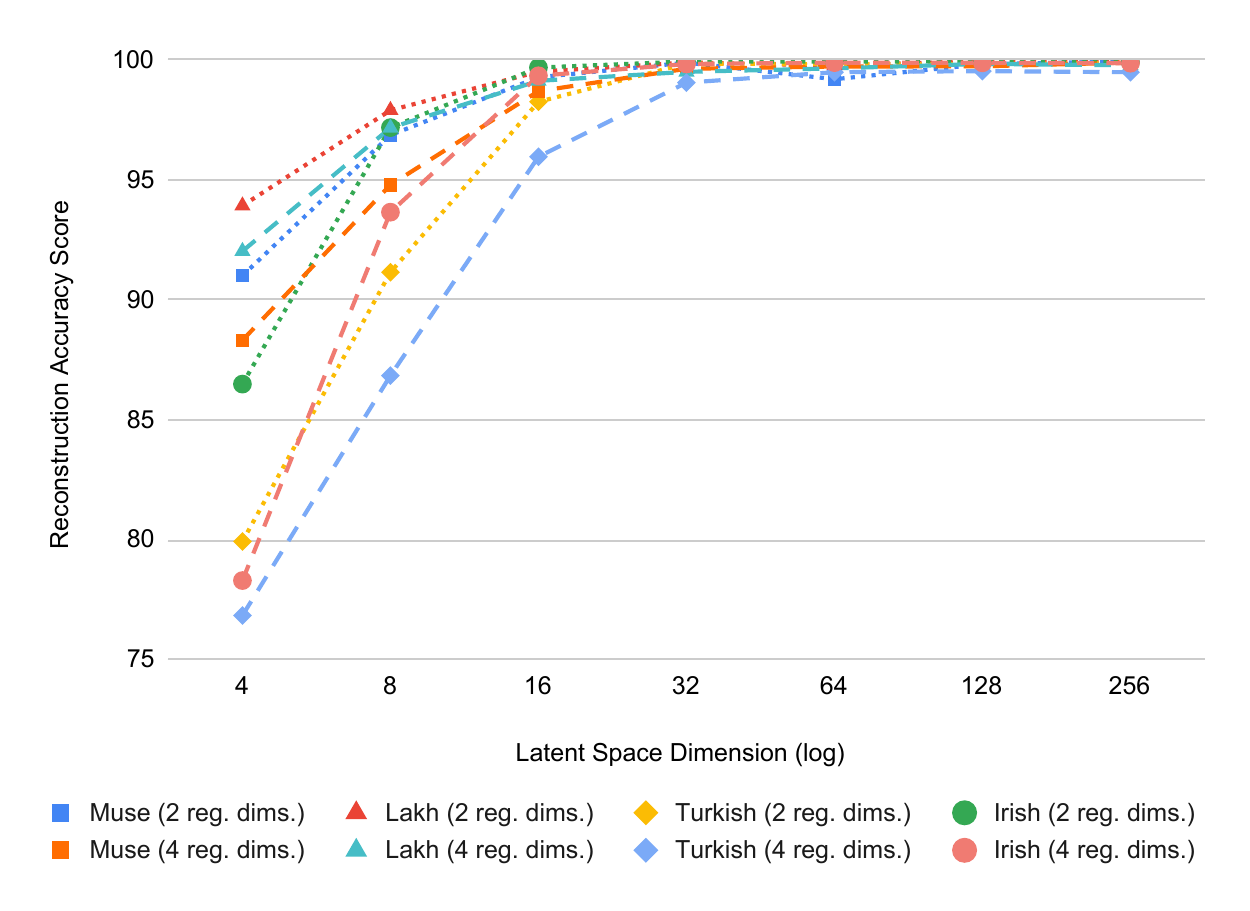}
    \caption{Reconstruction Accuracy for MeasureVAE}
    \label{fig:ReconstructionAccuracy}
\end{figure}

\begin{figure}
    \centering
    \includegraphics[width = 7.5cm]{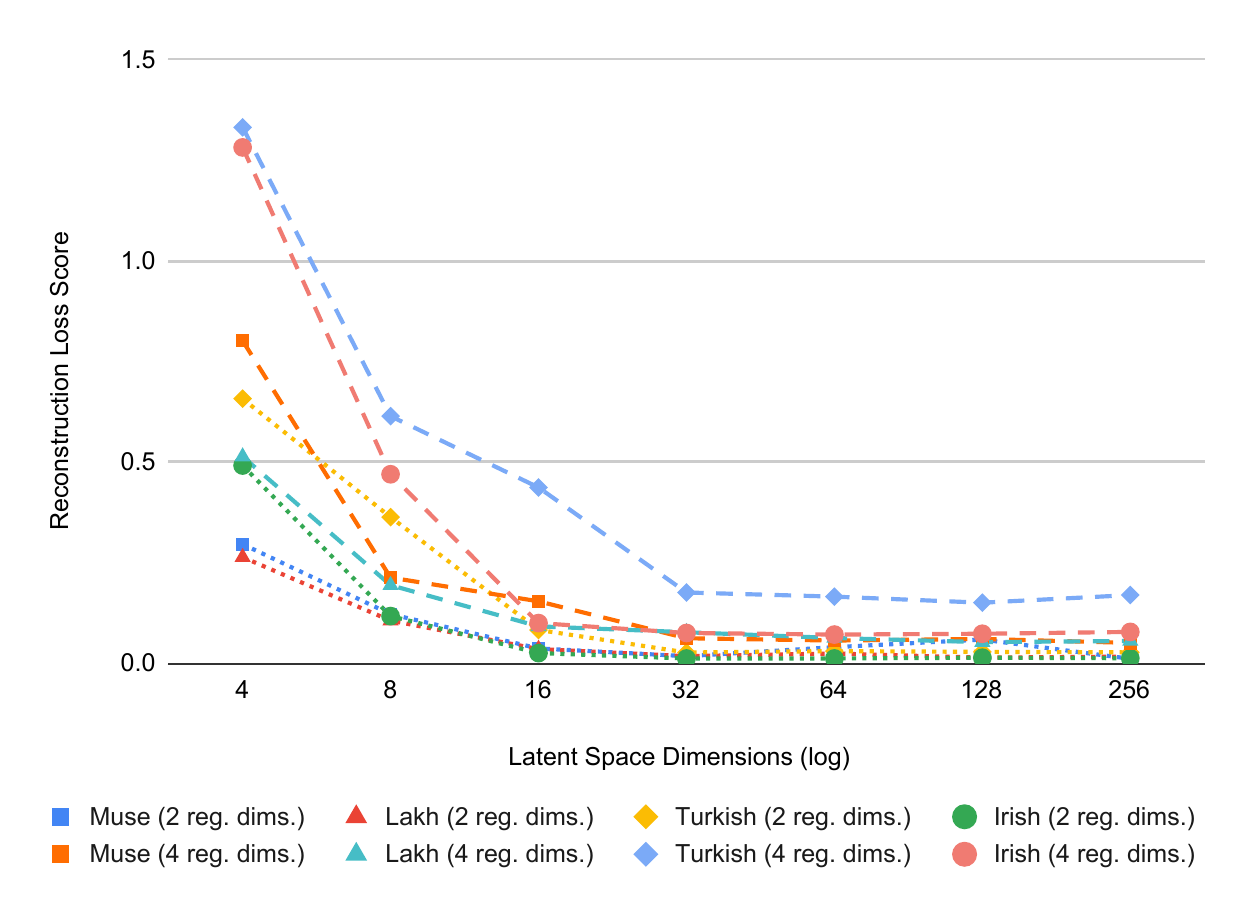}
    \caption{Loss for MeasureVAE}
    \label{fig:ReconstructionLoss}
\end{figure}

Fig. \ref{fig:ReconstructionAccuracy} and \ref{fig:ReconstructionLoss} illustrate the average Reconstruction Accuracy and Loss respectively for each dataset across all dimensions and with 2 (ND\&RC) or 4 regularised dimensions. 

The results indicate that MeasureVAE performed best with the Lakh Clean dataset with the lowest Loss scores and the highest Reconstruction Accuracy scores for both 2 and 4 regularised dimensions. This may be a result of the Lakh Clean dataset being less musically complex with the lowest mean Note Density, Note Range, and Rhythmic Complexity (Table \ref{tab:datasets}) arguably making the modelling task easier and \emph{suggesting that MeasureVAE may be more suited to less complex musical styles}. MeasureVAE performed least well for the Turkish Makam dataset which had the highest average Loss scores and lowest Reconstruction Accuracy scores. MeasureVAE's poor performance with the Turkish Makam dataset may be due to the higher complexity of the music in the dataset with high mean Note Density and Rhythmic Complexity, and lower Average Internal Jump than other datasets. Moreover, there are similar poor Reconstruction Accuracy and Loss scores found for the Irish Folk dataset which also has high mean ND, NR, and RC. The poor performance of the Turkish Makam dataset may also be  may also be due to it being the smallest of the datasets used in this experiment, or the complex tonal features of Turkish Makam music \cite{Dzhambazov2016OnTU} which may not be captured in the musical metrics used in this experiment.

The results in Table \ref{table:MeasureVAELossReconstruction} suggest that 2 regularised dimensions perform better than 4 for Loss and Reconstruction Accuracy scores. The pair of ND\&RC regularised dimensions performing better than NR\&AIJ for both Loss and Reconstruction Accuracy scores.
The results also indicate that Reconstruction Accuracy and Loss scores improve up to 32 dimensional latent space and plateau for larger latent spaces.

\subsubsection{Attribute Interpretability}

Table \ref{table:MeasureVAEInterpretability} shows that MeasureVAE is capable of generating music with musical attribute Interpretability scores of at least 0.8 for all of the tested musical attributes, though not for all latent space sizes. Interpretability scores for 4 regularised dimensions are over 0.95 for 69 of the possible 112 combinations suggesting that MeasureVAE is able to generate music with good musical interpretability with 4 regularised dimensions. With 2 regularised dimensions Interpretability scores are above 0.95 for Note Density, Rhythmic Complexity, and Note Range for all datasets and the majority of latent dimensions (89 of 98) except RC for the Irish Folk dataset which reached a maximum RC Interpretability of 0.832 and AIJ for the Lakh dataset (maximum 0.897).
Overall, this \emph{suggests that MeasureVAE can successfully generate music across Folk, Classical, and Popular music with interpretable musical dimensions of control}. The results also indicate that different datasets have different Interpretability scores for different musical attributes, though it is not possible at this stage to say whether these Interpretability scores are good nor not cf. \cite{adel_interpretability}.

\begin{figure}
    \centering
    \includegraphics[width = 7.5cm]{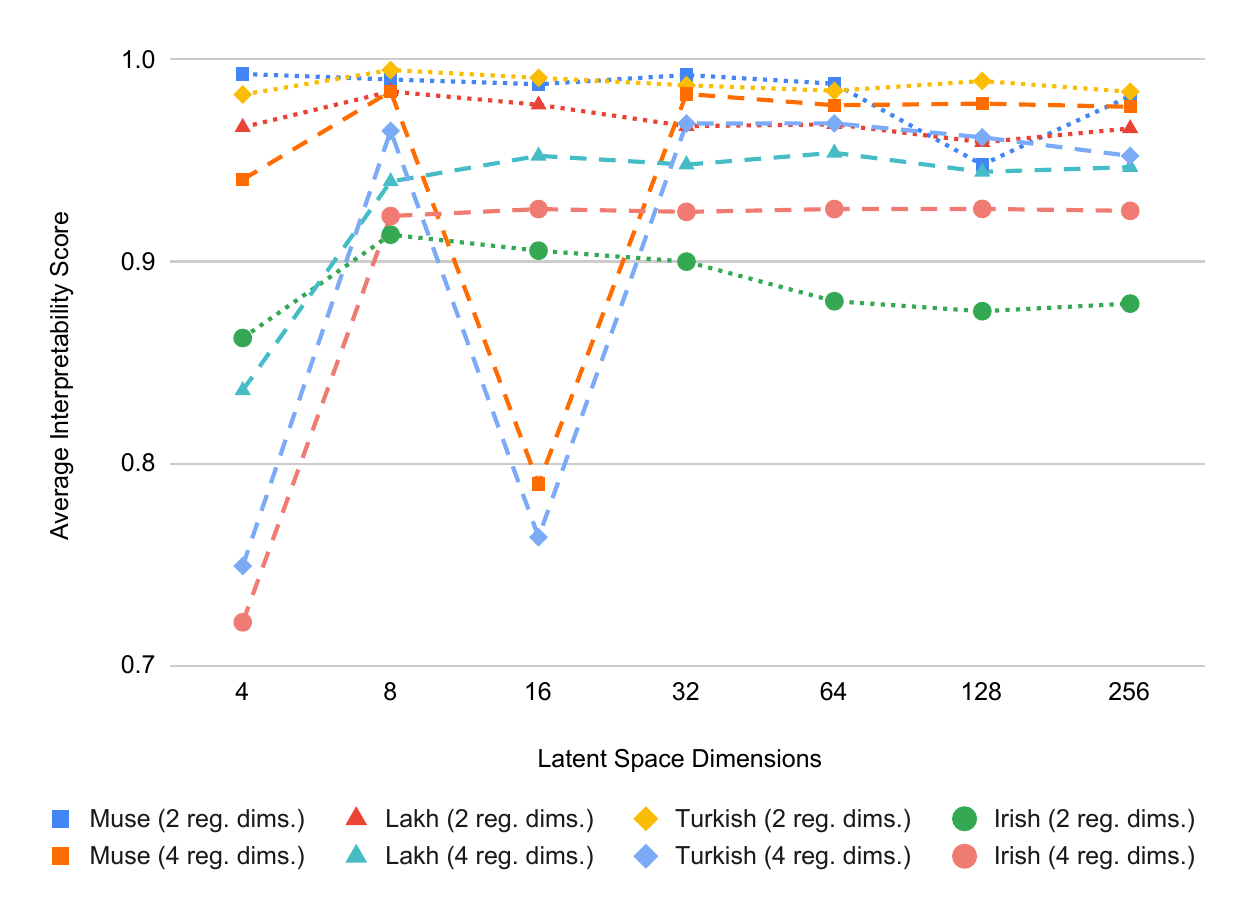}
    \caption{Average Interpretability for MeasureVAE Latent Regularised Dimensions.}
    \label{fig:AverageInterpretability}
\end{figure}

\textbf{Number of regularised dimensions.} For ease of inspection, Fig. \ref{fig:AverageInterpretability} illustrates the average Interpretability scores for all attributes for each dataset and 4 and 2 regularised dimensions (ND\&RC). 
As suggested in Fig. \ref{fig:AverageInterpretability} and detailed in Table \ref{table:MeasureVAEInterpretability}, Interpretability scores for each regularised attribute were in general higher for 2 regularised dimensions than 4 which is to be expected as it is easier to achieve successful and linearly independent regularisation in fewer dimensions. The exception to this are the Rhythmic Complexity Interpretability scores for Muse Bach, Lakh Clean, and Irish Folk datasets. For Muse Back and Lakh Clean datasets the mean RC Interpretability scores were equal. The Irish Folk dataset's mean Rhythmic Complexity Interpretability scores are marginally higher for 4 regularised dimensions than 2 regularised due to the performance for latent spaces of 64, 128, and 256 dimensions. Inspecting the Interpretability scores for each dimension the data suggests that 2 regularised dimensions perform best compared to 4 regularised dimensions for Note Density and Average Interval Jump Interpretability scores.
The higher mean Interpretability scores for 2 regularised dimensions versus 4 may be due to only regularising 2 dimensions rather than the nature of the regularisation itself. The poor performance of Rhythm Complexity for the Irish Folk dataset may be a reflection of higher Note Density, Note Range, and Rhythmic Complexity means for the Irish Folk dataset, or it may be a reflection of the larger dataset size.

\textbf{Interpretability score performance.} Rankings of Interpretability scores for datasets are not consistent across the number of dimensions in the latent space. For example, for 4 regularised dimensions the highest Rhythmic Complexity Interpretability score with 16 latent space dimensions is for the Lakh Clean dataset (0.974) whereas for 32 latent space dimensions the highest RC Interpretability score is for the Muse Bach dataset (0.985).
For 4 regularised dimensions the highest scoring attribute Interpretability scores are consistent for 64, 128, and 256 dimensions e.g. the Muse Bach dataset has the highest Note Range Interpretability scores for latent spaces with 64, 128, and 256 dimensions.
For 2 regularised dimensions there is no consistently highest ranked attribute for Interpretability across the range of latent space dimensions.

\textbf{Optimal configurations.} Given that high Reconstruction Accuracy scores and low Loss scores are reached at a 32 dimensional latent space for both 2 and 4 regularised dimensions, and that rankings of Interpretability scores stabilise at 64 dimensions and above, the \emph{results suggest that a 32 or 64 dimensional latent space would be when optimal applying MeasureVAE across genres} as it minimises latent space size and Loss whilst maximising Reconstruction Accuracy and providing similar Interpretability scores to higher dimensional spaces. However careful selection of latent space size is recommended when MeasureVAE is to be used to generate specific genres of music. For example, 16 or even 8 latent dimensions are likely to be optimal for Irish Folk music generation with 2 regularised dimensions given that its best Interpretability performance is with an 8 dimensional latent space.

\textbf{Dataset performance.} Taking the average Interpretability score across all latent space sizes, results suggest that for 4 regularised dimensions, the Irish Folk dataset has the highest average Interpretability scores for ND and NR, whereas Muse Bach has the highest for RC and AIJ. It is worth noting that the Irish Folk dataset itself has the highest mean ND and NR and the second highest mean RC (Table \ref{tab:datasets}), whereas the the Muse Bach dataset has the second lowest mean RC and lowest mean AIJ \emph{suggesting that there is not a correlation between the musical attributes of the datasets and the average Interpretability scores of the MeasureVAE models}. In terms of lowest average Interpretability scores, the Turkish Makam dataset has the lowest for ND, Irish Folk dataset the lowest for RC, the Muse Bach and Turkish Makam datasets equally have the lowest for NR, and the Lakh Clean dataset has the lowest average AIJ Interpretability score.

For 2 regularised dimensions, the Turkish Makam dataset has the highest average ND and RC Interpretability scores whereas Lakh Clean has the lowest ND, and Irish Folk dataset has the lowest RC. Note that the Irish folk dataset has the lowest Rhythmic Complexity Interpretability scores for both 2 and 4 regularised dimensions and for all sizes of latent spaces. Whilst the Irish Folk dataset has a high mean Rhythmic Complexity this does not explain the poor RC Interpretability score for the Irish Folk dataset as the Turkish Makam dataset has the highest mean RC and also highest RC Interpretability score for 2 regularised dimensions, \emph{suggesting that there is not a correlation between the musical attributes of the datasets and the Interpretability scores of the MeasureVAE models}.

Fig. \ref{fig:AverageInterpretability} and Table \ref{table:MeasureVAEInterpretability} indicate some anomalies in the Interpretability scores. For 2 regularised dimensions, there is an outlying Rhythmic Complexity Interpretability score for 128 latent dimensions. For 4 regularised dimensions in a 16 dimensional latent space there are outlying Interpretability scores the Muse Bach dataset (ND and NR) and the Turkish Makam dataset (ND and RC). On investigation of the data no obvious reasons for these outlying results emerge. Instead, these anomalies \emph{suggest potential inconsistent performance of MeasureVAE for different datasets and latent dimension size}, necessitating careful selection of latent space size for musical style.

\section{Conclusions}

\rr{This is the first time that two VAE models with semantic features for control of music generation have been systematically compared in terms of performance, latent space features, musical attributes, and training datasets. In doing this we help researchers to understand the properties of state-of-the-art generative models and so help to inform generative model research and design by providing a detailed analysis of current systems.}
We found that MeasureVAE has higher Reconstruction Accuracy and Reconstruction Efficiency than AdversarialVAE but lower musical Attribute Independence (Section \ref{sec:compareVAEmodels}). 

The results also show that MeasureVAE is capable of generating music across folk, pop, rock, jazz and blues, R\&B, and classical music styles, and performs best with lower complexity musical styles such as pop and rock. Furthermore, results show that MeasureVAE was able to generate music across these genres with interpretable musical dimensions of control.

The MeasureVAE generated output was found to have different musical Interpretability scores for different datasets, but there was not a correlation between the musical features of datasets and the related Interpretability scores of the generated music. 
For 4 regularised dimensions, the Irish Folk dataset has the highest average Interpretability scores for Note Density and Note Range, whereas Muse Bach has the highest for Rhythmic Complexity and Average Interval Jump Interpretability scores. 

Interpretability metrics were in general higher when only two dimensions of the latent space were regularised. Similarly, Loss and Reconstruction Accuracy scores were better fro two regularised dimensions than four. These findings are to be expected as it is easier to achieve successful and linearly independent regularisation in fewer dimensions. For Loss and Reconstruction Accuracy scores, MeasureVAE performed better with the pair of Note Density and Rhythmic Complexity regularised dimensions than when trained with Note Range and Average Interval Jump regularised dimensions. This may be because MeasureVAE is better at generating the tonal and rhythmic aspects of the music which are captured by ND and RC.

In terms of recommendations for use, results suggest that a 32 or 64 dimensional latent space would be optimal using MeasureVAE to generate music across a range of genres as this minimises latent space size whilst maximising reconstruction performance and providing similar Interpretability scores to those offered by higher dimensional spaces. However, careful selection of latent space size is required for generation of specific genres of music. For example, Irish Folk music may be optimally generated with 16 or even 8 latent dimensional space.

These results show that when explainable features are added to the MeasureVAE system it performs well across genres of musical generation. For XAI and the Arts more broadly our approach demonstrates how complex AI models can be compared and explored in order to identify optimal configurations for the required styles of music generation. The work also demonstrates the complex relationships between datasets, explainable attributes, and AI model music generation performance.
\rr{This complex relationship has some wider implications for generative AI models. For example, it highlights the bias built in to models which makes them more amenable to certain datasets rather than others - a key concern of Human-Centred AI \cite{Garibay2023}. In our case the structure of MeasureVAE biased it towards lower complexity musical styles such as pop and rock at the expense of more complex and forms of music such as Turkish Makram which it is worth noting are more marginalised forms of music.}

The research presented here is a first step and is limited in scope. Future research needs to explore the effect that other genres and datasets, dataset sizes, musical attributes, and training regimes have on the performance of explainable AI models. \rr{This would provide a more in-depth analysis of the landscape of generative AI models from which to inform future AI model research and design.} For example, we chose two sets of musical attributes to use in this experiment based on frequently used attributes in research papers but the utility of musical attributes to control musical generation very much depends on the context of use. We also need to compare a wider range of generative models and explainability techniques across datasets and musical attributes to identify which combinations of explainable AI model and dataset offer the best generative performance for the musical features of interest to musicians. \rr{For example, using information dynamic measures to compare generative models following \cite{Koh2018}. It would also be important to examine longer-term music generation such as song structure generation e.g. \cite{Chen2020}, and to use subjective listening tests to better understand the quality of the music generated (ibid.).}
\rr{Exploring how the robustness and interpretability of the models tested could be improved, for example \cite{Liu2022}, would be especially important for real-time music generation settings such as live performance.}
\rr{Moreover, it would be useful more broadly to explore how the evaluation approach deployed in this paper could be applied to other domains such as image generation. For example, how comparative evaluations of image generating VAEs \cite{Wei2020} could be undertaken to compare interpretable features as we have done in this paper, or to apply our approach to comparing the effect of different interpretable features on the robustness of image generation \cite{Liu2022} instead of music generation.}

Finally, we need to start to explore how the explainable features of the models tested in this paper could be used to make more interactive generative systems that move beyond being a empirical tool for researching AI models to become more of a creative tool used in musical practice and performance. As a first step we will build the findings of this research into audio plugins which can be embedded into musician's musical tool chains as part of their artistic practice, starting with a MIDI music processor \cite{AAAI23}.

\section*{Acknowledgments}

This work was supported by the UKRI Centre for Doctoral Training in Artificial Intelligence and Music supported by UKRI (EP/S022694/1), Queen Mary University of London, and the Carleton College Career Center for funding.

This paper includes data from research undertaken as part of Zhang's Masters project and Zhao's Summer internship at Queen Mary University of London.

\section*{Author Contributions}

\rr{Bryan-Kinns instigated this work, led the research, supervised the student projects, and led the data analysis and writing. Zhao and Zhang contributed equally to the implementation of AI models, data collection and analysis in this work. Banar developed the original implementation in \cite{XAIMS_NeurIps2021} which formed the basis for this work, contributed to the supervision of the student projects, contributed to the data analysis, and led the technical writing in this paper.}

\bibliographystyle{unsrt} % unsorted bib (i.e. by appearance)

%\bibliography{references}% common bib file

%% %%%%%%%%%%%%%%%%%%%%%%%%%%%%%%%%%%%

% Biography see template

%\begin{figure}[h]%
%\centering
%\includegraphics[width=0.3\textwidth]{MIR Latex Template/Photo_NickBryan-Kinns.png}
%\caption{Second author}%\label{fig1}
%\end{figure}

%\end{biography}

%%%%%% big tables %%%%%%%%%%%%%%%%%%%%%%%%%%%%%%%%%

%\include{MIR Latex Template/TableLossReconstruction}

\begin{table*}[h]																	
\begin{center}																	
\begin{minipage}{\textwidth}																	
\caption{Loss and Reconstruction Accuracy scores for MeasureVAE.}\label{table:MeasureVAELossReconstruction}																	
\begin{tabular*}{\textwidth}{@{\extracolsep{\fill}}c|cccc|cccc@{\extracolsep{\fill}}}																	
%\begin{tabular}{@{}ccccc@{}}																	
\toprule%																	
{Latent}	&	\multicolumn{4}{c|}{{Loss Scores}}							&	\multicolumn{4}{c}{{Reconstruction Accuracy Scores (\%)}}							\\
{Dimensions}	&	Muse	&	Lakh	&	Turkish	&	Irish	&	Muse	&	Lakh	&	Turkish	&	Irish	\\
\midrule																	
	&	\multicolumn{8}{c}{{2 Regularised Dimensions (ND\&RC)}}															\\
\midrule																	
4	&	0.296	&	0.264	&	0.658	&	0.492	&	91.009	&	93.928	&	79.925	&	86.488	\\
8	&	0.122	&	0.106	&	0.364	&	0.118	&	96.873	&	97.909	&	91.151	&	97.185	\\
16	&	0.037	&	0.036	&	0.083	&	0.026	&	99.284	&	99.517	&	98.260	&	99.688	\\
32	&	0.018	&	0.018	&	0.027	&	0.012	&	99.894	&	99.900	&	99.853	&	99.942	\\
64	&	0.040	&	0.025	&	0.031	&	0.012	&	99.203	&	99.777	&	99.827	&	99.939	\\
128	&	0.059	&	0.014	&	0.029	&	0.014	&	99.828	&	99.958	&	99.837	&	99.951	\\
256	&	0.011	&	0.017	&	0.027	&	0.013	&	99.947	&	99.920	&	99.879	&	99.912	\\
\textit{Mean	} & \textit{	0.084	} & \textit{	0.068	} & \textit{	0.174	} & \textit{	0.098	} & \textit{	98.005	} & \textit{	98.701	} & \textit{	95.533	} & \textit{	97.586	} \\
																	
\midrule																	
	&	\multicolumn{8}{c}{{2 Regularised Dimensions (NR\&AIJ)}}															\\
\midrule																	
4	&	0.372	&	0.265	&	0.823	&	0.706	&	89.972	&	94.162	&	75.627	&	82.661	\\
8	&	0.163	&	0.145	&	0.512	&	0.190	&	95.788	&	97.924	&	89.187	&	96.552	\\
16	&	0.088	&	0.067	&	0.256	&	0.083	&	98.532	&	99.109	&	96.168	&	99.389	\\
32	&	0.049	&	0.116	&	0.137	&	0.058	&	99.649	&	98.340	&	99.181	&	99.796	\\
64	&	0.041	&	0.047	&	0.128	&	0.055	&	99.799	&	99.662	&	99.496	&	99.841	\\
128	&	0.042	&	0.029	&	0.146	&	0.054	&	99.764	&	99.909	&	99.323	&	99.876	\\
256	&	0.040	&	0.033	&	0.136	&	0.052	&	99.806	&	99.893	&	99.448	&	99.870	\\
\textit{Mean	} & \textit{	0.114	} & \textit{	0.100	} & \textit{	0.306	} & \textit{	0.171	} & \textit{	97.616	} & \textit{	98.428	} & \textit{	94.061	} & \textit{	96.855	} \\
\midrule																	
	&	\multicolumn{8}{c}{{4 Regularised Dimensions}}															\\
\midrule																	
4	&	0.803	&	0.514	&	1.333	&	1.283	&	88.317	&	92.031	&	76.837	&	78.294	\\
8	&	0.213	&	0.194	&	0.615	&	0.470	&	94.796	&	97.176	&	86.843	&	93.658	\\
16	&	0.154	&	0.092	&	0.438	&	0.100	&	98.705	&	99.135	&	95.967	&	99.346	\\
32	&	0.063	&	0.077	&	0.176	&	0.076	&	99.638	&	99.510	&	99.061	&	99.838	\\
64	&	0.057	&	0.063	&	0.166	&	0.072	&	99.733	&	99.656	&	99.495	&	99.894	\\
128	&	0.061	&	0.053	&	0.151	&	0.074	&	99.740	&	99.844	&	99.540	&	99.886	\\
256	&	0.051	&	0.056	&	0.170	&	0.078	&	99.900	&	99.781	&	99.495	&	99.871	\\
\textit{Mean	} & \textit{	0.200	} & \textit{	0.150	} & \textit{	0.436	} & \textit{	0.308	} & \textit{	97.261	} & \textit{	98.162	} & \textit{	93.891	} & \textit{	95.827	} \\
\botrule																	
																	
\end{tabular*}																	
%\end{tabular}																	
\end{minipage}																	
\end{center}																	
\end{table*}																	

% \include{MIR Latex Template/TableInterpretability}
%\note{REMEMBER TO INCLUDE THE TABLE TEX directly IN THE TEX DOCUMENT}

\begin{table*}[h]																	
\begin{center}																	
\begin{minipage}{\textwidth}																	
\caption{Interpretability scores for MeasureVAE. Bold indicates the highest score of 2 and 4 dimensions.}\label{table:MeasureVAEInterpretability}																	
\begin{tabular*}{\textwidth}{@{\extracolsep{\fill}}c|cccc|cccc@{\extracolsep{\fill}}}																	
\toprule%																	
																	
{Latent}	&	\multicolumn{4}{c|}{{}}							&	\multicolumn{4}{c}{{}}							\\
{Dimensions}	&	Muse	&	Lakh	&	Turkish	&	Irish	&	Muse	&	Lakh	&	Turkish	&	Irish	\\
\midrule																	
	&	\multicolumn{4}{c|}{{2 Regularised Dimensions (ND\&RC)}}							&	\multicolumn{4}{c}{{4 Regularised Dimensions}}							\\
\midrule																	
	&	\multicolumn{8}{c}{{Note Density (ND) Interpretability Scores}}															\\
\midrule																	
{4	} & \textbf{	0.999	} & \textbf{	0.978	} & \textbf{	0.996	} & \textbf{	0.992	} & {	0.982	} & {	0.956	} & {	0.940	} & {	0.931	} \\
{8	} & {	0.996	} & \textbf{	0.992	} & \textbf{	1.000	} & \textbf{	0.996	} & \textbf{	0.998	} & {	0.977	} & {	0.998	} & {	0.991	} \\
{16	} & \textbf{	0.997	} & \textbf{	0.990	} & \textbf{	0.999	} & \textbf{	0.997	} & {	0.708	} & {	0.986	} & {	0.584	} & {	0.996	} \\
{32	} & \textbf{	0.996	} & \textbf{	0.983	} & \textbf{	0.998	} & {	0.994	} & {	0.992	} & {	0.975	} & {	0.994	} & \textbf{	0.995	} \\
{64	} & \textbf{	0.996	} & \textbf{	0.977	} & \textbf{	0.997	} & {	0.992	} & {	0.990	} & {	0.975	} & {	0.994	} & \textbf{	0.995	} \\
{128	} & {	0.987	} & \textbf{	0.981	} & \textbf{	0.998	} & {	0.990	} & \textbf{	0.989	} & {	0.969	} & {	0.994	} & \textbf{	0.997	} \\
{256	} & \textbf{	0.990	} & \textbf{	0.978	} & \textbf{	0.999	} & {	0.989	} & {	0.989	} & {	0.977	} & {	0.994	} & \textbf{	0.994	} \\
\textit{{Mean	}} & \textbf{\textit{	0.995	}} & \textbf{\textit{	0.983	}} & \textbf{\textit{	0.998	}} & \textbf{\textit{	0.993	}} & {\textit{	0.950	}} & {\textit{	0.974	}} & {\textit{	0.928	}} & {\textit{	0.985	}} \\
\midrule																	
	&	\multicolumn{8}{c}{{Rhythmic Complexity (RC) Interpretability Scores}}															\\
\midrule																	
{4	} & \textbf{	0.987	} & \textbf{	0.956	} & \textbf{	0.970	} & \textbf{	0.732	} & {	0.946	} & {	0.927	} & {	0.925	} & {	0.712	} \\
{8	} & {	0.985	} & \textbf{	0.977	} & \textbf{	0.991	} & \textbf{	0.832	} & \textbf{	0.989	} & {	0.959	} & {	0.989	} & {	0.774	} \\
{16	} & \textbf{	0.980	} & {	0.966	} & \textbf{	0.983	} & \textbf{	0.815	} & {	0.955	} & \textbf{	0.974	} & {	0.580	} & {	0.802	} \\
{32	} & \textbf{	0.989	} & {	0.952	} & {	0.977	} & \textbf{	0.806	} & {	0.985	} & \textbf{	0.965	} & {	0.977	} & {	0.800	} \\
{64	} & \textbf{	0.981	} & {	0.960	} & \textbf{	0.973	} & {	0.769	} & {	0.975	} & \textbf{	0.967	} & {	0.971	} & \textbf{	0.801	} \\
{128	} & {	0.909	} & {	0.937	} & \textbf{	0.981	} & {	0.761	} & \textbf{	0.980	} & \textbf{	0.945	} & {	0.976	} & \textbf{	0.800	} \\
{256	} & {	0.975	} & {	0.955	} & {	0.970	} & {	0.769	} & \textbf{	0.976	} & \textbf{	0.961	} & \textbf{	0.975	} & \textbf{	0.802	} \\
\textit{{Mean	}} & {\textit{	0.972	}} & {\textit{	0.957	}} & \textbf{\textit{	0.978	}} & {\textit{	0.783	}} & {\textit{	0.972	}} & {\textit{	0.957	}} & {\textit{	0.913	}} & \textbf{\textit{	0.784	}} \\
\toprule																	
	&	\multicolumn{4}{c|}{{2 Regularised Dimensions (NR\&AIJ)}}							&	\multicolumn{4}{c}{{4 Regularised Dimensions}}							\\
%{Dimensions}	&	Muse	&	Lakh	&	Turkish	&	Irish	&	Muse	&	Lakh	&	Turkish	&	Irish	\\
\midrule																	
	&	\multicolumn{8}{c}{{Note Range (NR) Interpretability Scores}}															\\
\midrule																	
{4	} & \textbf{	0.961	} & \textbf{	0.864	} & \textbf{	0.953	} & \textbf{	0.950	} & {	0.922	} & {	0.811	} & {	0.460	} & {	0.736	} \\
{8	} & \textbf{	0.984	} & {	0.950	} & {	0.993	} & {	0.955	} & {	0.981	} & \textbf{	0.961	} & {	0.993	} & \textbf{	0.991	} \\
{16	} & \textbf{	0.975	} & {	0.958	} & {	0.945	} & {	0.963	} & {	0.540	} & \textbf{	0.971	} & \textbf{	0.974	} & \textbf{	0.993	} \\
{32	} & {	0.977	} & \textbf{	0.969	} & {	0.976	} & \textbf{	0.990	} & \textbf{	0.983	} & {	0.966	} & \textbf{	0.988	} & {	0.987	} \\
{64	} & {	0.975	} & {	0.965	} & \textbf{	0.983	} & {	0.988	} & \textbf{	0.976	} & \textbf{	0.975	} & {	0.982	} & {	0.988	} \\
{128	} & {	0.971	} & {	0.968	} & {	0.970	} & {	0.989	} & \textbf{	0.976	} & \textbf{	0.971	} & \textbf{	0.980	} & \textbf{	0.991	} \\
{256	} & {	0.974	} & {	0.968	} & \textbf{	0.983	} & {	0.957	} & \textbf{	0.978	} & \textbf{	0.971	} & {	0.977	} & \textbf{	0.990	} \\
\textit{{Mean	}} & \textbf{\textit{	0.974	}} & \textbf{\textit{	0.949	}} & \textbf{\textit{	0.972	}} & \textbf{\textit{	0.970	}} & {\textit{	0.908	}} & {\textit{	0.947	}} & {\textit{	0.908	}} & {\textit{	0.954	}} \\
\midrule																	
	&	\multicolumn{8}{c}{{Average Interval Jump (AIJ) Interpretability Scores}}															\\
\midrule																	
{4	} & \textbf{	0.949	} & \textbf{	0.714	} & {	0.621	} & \textbf{	0.844	} & {	0.914	} & {	0.651	} & \textbf{	0.672	} & {	0.506	} \\
{8	} & {	0.960	} & \textbf{	0.869	} & \textbf{	0.929	} & \textbf{	0.964	} & \textbf{	0.970	} & {	0.863	} & {	0.881	} & {	0.934	} \\
{16	} & \textbf{	0.976	} & {	0.868	} & {	0.886	} & \textbf{	0.973	} & {	0.957	} & \textbf{	0.880	} & \textbf{	0.916	} & {	0.915	} \\
{32	} & \textbf{	0.981	} & \textbf{	0.896	} & \textbf{	0.951	} & \textbf{	0.969	} & {	0.974	} & {	0.888	} & {	0.916	} & {	0.918	} \\
{64	} & {	0.967	} & {	0.882	} & \textbf{	0.958	} & \textbf{	0.970	} & \textbf{	0.970	} & \textbf{	0.899	} & {	0.928	} & {	0.922	} \\
{128	} & \textbf{	0.974	} & \textbf{	0.897	} & {	0.887	} & \textbf{	0.971	} & {	0.969	} & {	0.894	} & \textbf{	0.898	} & {	0.918	} \\
{256	} & \textbf{	0.976	} & \textbf{	0.897	} & \textbf{	0.894	} & \textbf{	0.968	} & {	0.964	} & {	0.879	} & {	0.865	} & {	0.916	} \\
\textit{{Mean	}} & \textbf{\textit{	0.969	}} & \textbf{\textit{	0.860	}} & \textbf{\textit{	0.875	}} & \textbf{\textit{	0.951	}} & {\textit{	0.960	}} & {\textit{	0.851	}} & {\textit{	0.868	}} & {\textit{	0.861	}} \\
\botrule																	
\end{tabular*}																	
																	
\end{minipage}																	
\end{center}																	
\end{table*}

%%%%%% big tables %%%%%%%%%%%%%%%%%%%%%%%%%%%%%%%%%

\end{document}